\documentclass[aps,preprint,showpacs,epsfig] {revtex4}
\usepackage{pdfsync}
\usepackage{graphicx}  
\usepackage{epstopdf}
\usepackage{amsmath}

\begin{document}

\title{Modeling the Field Control of the Surface Electroclinic Effect near Continuous and First Order Smectic-$A^*$-Smectic-$C^*$ Transitions}

\author{Kara Zappitelli, Dana N. Hipolite, and Karl Saunders}

\affiliation{Department of Physics, California Polytechnic State
University, San Luis Obispo, CA 93407, USA}

\email{ksaunder@calpoly.edu}

\date{\today}

% Last modified 11/6/13

\begin{abstract}

We present and analyze a model for the combination of bulk and surface electroclinic effects in the smectic-$A^*$ (Sm-$A^*$) phase near a Sm-$A^*$--Sm-$C^*$ transition. As part of our analysis we calculate the dependence of the surface tilt on external electric field and show that it can be eliminated, or even reversed from its zero-field value. This is in good agreement with previous experimental work on a system (W415) with a continuous Sm-$A^*$--Sm-$C^*$ transition. We also analyze, for the first time, the combination of bulk and surface electroclinic effects in systems with a first order Sm-$A^*$--Sm-$C^*$ transition. The variation of surface tilt with electric field in this case is much more dramatic, with discontinuities and hysteresis. With regard to technological, e.g., display, applications, this could be a feature to be avoided or potentially exploited. Near each type of Sm-$A^*$--Sm-$C^*$ transition we obtain the temperature dependence of the field required to eliminate surface tilt. Additionally, we analyze the effect of varying the system's enantiomeric excess, showing that it  strongly affects the field dependence of surface tilt, in particular, near a first order Sm-$A^*$--Sm-$C^*$ transition. In this case, increasing enantiomeric excess can change the field dependence of surface tilt from continuous to discontinuous. Our model also allows us to calculate the variation of layer spacing in going from surface to bulk, which in turn allows us to estimate the strain resulting from the difference between the surface and bulk layer spacing. We show that for certain ranges of applied electric field, this strain can result in layer buckling which reduces the overall quality of the liquid crystal cell. For de Vries materials, with small tilt-induced change in layer spacing, the induced strain for a given surface tilt should be smaller. However, we argue that this may be offset by the fact that de Vries materials, which typically have Sm-$A^*$--Sm-$C^*$ transitions near a tricritical point, will generally have larger surface tilt. 

\end{abstract}


\maketitle

\section{Introduction}
\label{Introduction}

\begin{figure}
\begin{center}
\includegraphics[scale=0.8]{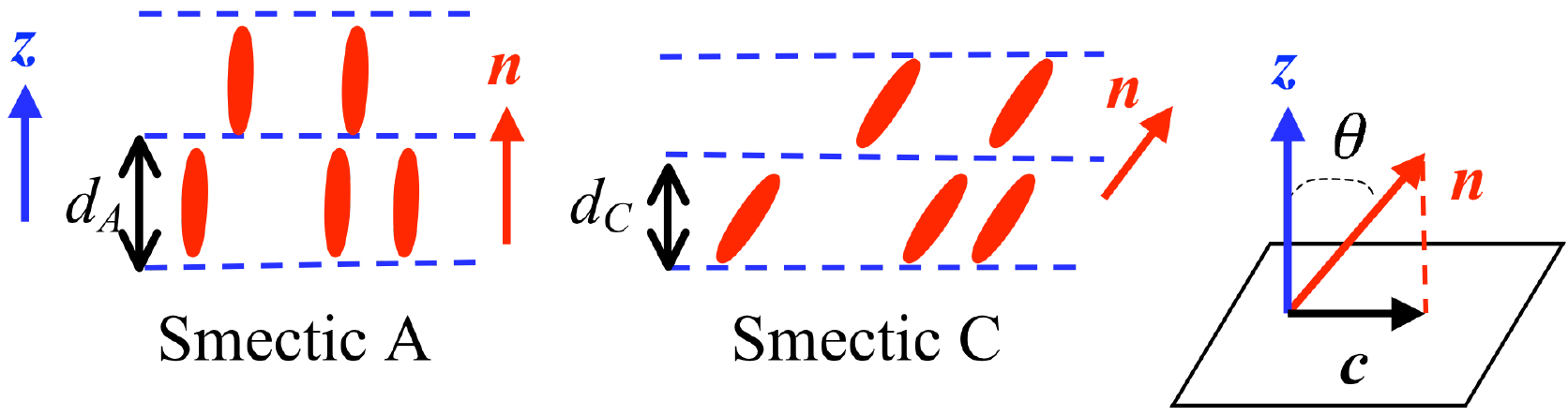}
\caption{A schematic view of the Sm-$A$ and Sm-$C$ phases. The transition from Sm-$A$ to Sm-$C$ phase can occur via a lowering of temperature, or in the case of a chiral, ferroelectric material, through the application of an electric field, i.e., via the bulk electroclinic effect. In the latter case, the electric field would point perpendicular to both the layer normal ${\hat z}$ and the azimuthal direction ${\bf c}$, i.e, into/out of the page.}
\label{smectic cartoon}
\end{center}
\end{figure}

Much work has been done over the past four decades to investigate smectic-$A$ (Sm-$A$) and smectic-$C$ (Sm-$C$) liquid crystal phases, particularly the phase transition between them. It is worthwhile briefly reviewing the basics of the non-chiral Sm-$A$ and Sm-$C$ phases. A key feature is that their density is modulated along one direction, taken to be ${\bf \hat z}$. This density modulation is often thought of in terms of the constituent molecules forming a layered structure, as shown schematically in Fig.~\ref{smectic cartoon}. Within the layers there is no long range positional order. In the Sm-$A$ phase the elongated molecules tend, on average, to align their long axes along a common direction ${\bf \hat n}$ (known as the director or the optical axis) that points along ${\bf \hat z}$, while in the Sm-$C$ phase ${\bf \hat n}$ lies at an angle $\theta$ relative to ${\bf \hat z}$. Figure~\ref{smectic cartoon} also shows the traditional view of the transition from Sm-$A$ to Sm-$C$: a tilting of the constituent molecules, and hence ${\bf \hat n}$, by an angle $\theta$ away from ${\bf \hat z}$. The molecules tilt in the same azimuthal direction ${\bf c}$, which is defined as the projection of ${\bf \hat n}$ onto the plane of the layers. The Sm-$A$ to Sm-$C$ phase transition will occur as temperature ($T$) is lowered through the transition temperature. As a result of the tilting of the molecules, the Sm-$C$ layer spacing, $d_C$, will be smaller than the Sm-$A$ layer spacing $d_A$, i.e., $d_C = d_A \cos(R\theta)$, where the reduction factor $R$ is a measure of the de Vries-like nature of the smectic \cite{LemieuxR}. For an ideal de Vries smectic, which exhibits no change in layer spacing at the transition, $R=0$, while in the traditional view of the transition, shown in Fig.~\ref{smectic cartoon}, $R\approx 1$.

More recently, considerable attention has been devoted to chiral smectics and the chiral Sm-$A^*$ -- Sm-$C^*$ transition (with the $^*$ denoting a chiral phase). Within the Sm-$C^*$ phase, the combination of tilt and chirality leads to a polarization within the layers which is perpendicular to ${\bf c}$, hence the use of the term ``ferroelectric" in describing chiral smectics. The ferroelectric nature of chiral smectics allows fast switching of the optical axis ${\bf \hat n}$ relative to the layer normal (i.e., a switching of tilt $\theta$) through the application of an electric field, perpendicular to both the layer normal ${\bf \hat z}$ and the azimuthal direction ${\bf c}$. If applied when the system is in the Sm-$A^*$ phase, the field induces a tilt of the molecules, i.e., it causes the system to transition to the Sm-$C^*$ phase. This phenomenon, known as the bulk electroclinic effect (BECE), was first predicted using a symmetry based argument \cite{Meyer} and was then observed experimentally \cite{Garoff and Meyer}. The BECE has led to considerable technological interest, particularly from the display industry, which in turn has prompted the synthesis of many new ferroelectric liquid crystal compounds.

\begin{figure}
\includegraphics[scale=1]{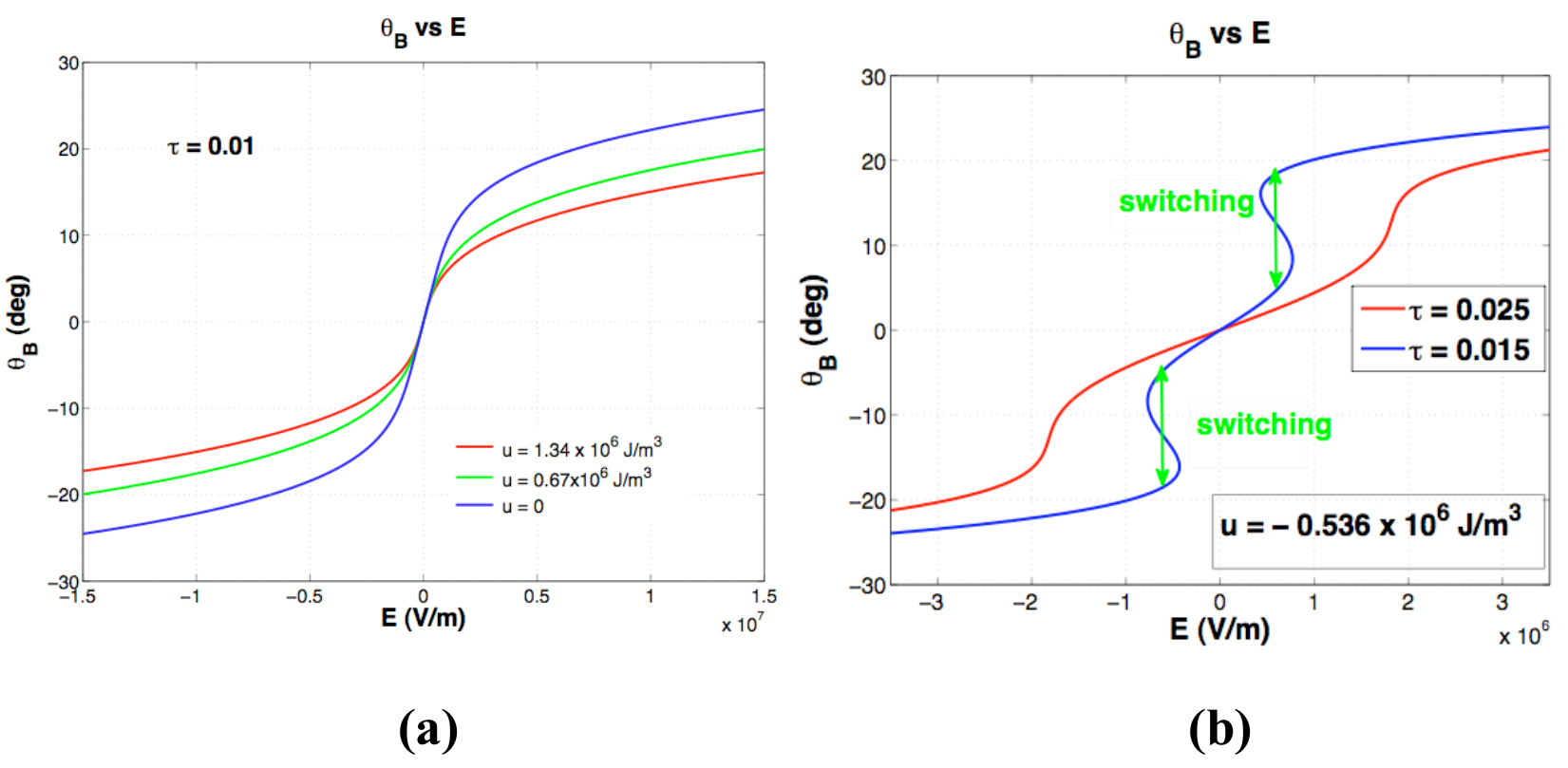}
\caption{Different types of electroclinic response curves $\theta_B(E)$ obtained using the theoretical model described in Section \ref{Bulk Electroclinic Effect}. (a) $\theta_B(E)$ curves for materials in the Sm-$A^*$ phase at $1\%$ reduced temperature (i.e., $\tau = (T-T_{AC})/T_{AC})=0.01$) above a continuous Sm-$A^*$ -- Sm-$C^*$ transition. Each of the three curves corresponds to a material with a transition at a different proximity to a tricritical point (TCP), with upper/lower (blue/red) being the closest to/furthest from the TCP, respectively. The parameter $u$ is a quantitative measure of this proximity and is defined via Eq.~(\ref{f_u}). One can see that the BECE is significantly enhanced with proximity to the TCP. (b) $\theta_B(E)$ for a material in the Sm-$A^*$ phase above a first order Sm-A$^*$ -- Sm-C$^*$ transition, i.e., with $u<0$. The lower (red) curve shows the response for $T>T_c$ (i.e., $\tau>\tau_{cb}$), with $\tau_{cb}$ a critical reduced temperature for response in the bulk. In this case the response is continuous but ``superlinear'', corresponding to positive curvature at small fields followed by negative curvature at large fields. The susceptibility $\frac{d\theta_B}{dE}$ is largest at the field where the curvature changes sign. As $T$ is lowered towards $T_c$ this value of susceptibility diverges. The upper (blue) curve shows the response for $T<T_c$ (i.e., $\tau<\tau_{cb}$). Now the response is discontinuous, leading to switching and possible hysteresis.}
\label{BECE_Response_curves}
\end{figure}

There are some details of the BECE in the Sm-$A^*$ phase that merit further discussion. An important characterization of the BECE is the electro-optical response curve $\theta_B(E)$, where $\theta_B$ is the tilt of the bulk optical axis and $E$ is the strength of the applied electric field. Different types of response curves (obtained using the model to be introduced in Section \ref{Bulk Electroclinic Effect}) are shown in Fig.~\ref{BECE_Response_curves}.  As shown in Fig.~\ref{BECE_Response_curves}(a), for systems with a continuous Sm-$A^*$ -- Sm-$C^*$ transition $\theta_B(E)$ is also continuous and the susceptibility $\frac{d\theta_B}{dE}$ is largest at $E=0$. The zero-field susceptibility diverges as the temperature $T$ is lowered towards the Sm-$A^*$ -- Sm-$C^*$ transition temperature, $T_{_{AC}}$. By now it is well established that many continuous Sm-$A^*$ -- Sm-$C^*$ transitions occur near or at a tricritical point \cite{Huang&Viner, HuangTP,ShashidharTP}. de Vries materials, in particular, seem to have transitions close to tricriticality \cite{LemieuxR}. At a tricritical point the nature of the transition changes from continuous to first order. For a continuous transition occurring near or at a tricritical point, the BECE is significantly enhanced, as shown in Fig.~\ref{BECE_Response_curves}(a). 

For systems with a first order Sm-$A^*$ -- Sm-$C^*$ transition the situation is quite different. Experimental and theoretical work \cite{Bahr&Heppke1,Bahr&Heppke2} on the BECE in materials with first order Sm-$A^*$ -- Sm-$C^*$ transitions demonstrated the existence of a critical point (in field ($E$) -- temperature ($T$) parameter space) which terminates a line of first order Sm-$C^*$ -- Sm-$C^*$ transitions. For temperatures above the critical temperature $T_c$ the response is continuous but exhibits what has been termed ``superlinear growth". As shown in Fig.~\ref{BECE_Response_curves}(b), this corresponds to positive curvature at small fields followed by negative curvature at large fields. It can also be seen that the susceptibility $\frac{d\theta_B}{dE}$ is largest at the field where the curvature changes sign. As $T$ is reduced towards $T_c$ this value of susceptibility diverges. For $T<T_c$ the response becomes discontinuous, as shown in Fig.~\ref{BECE_Response_curves}(b), leading to switching (at finite electric field values) and possible hysteresis. 

\begin{figure}
\begin{center}
\includegraphics[scale=0.6]{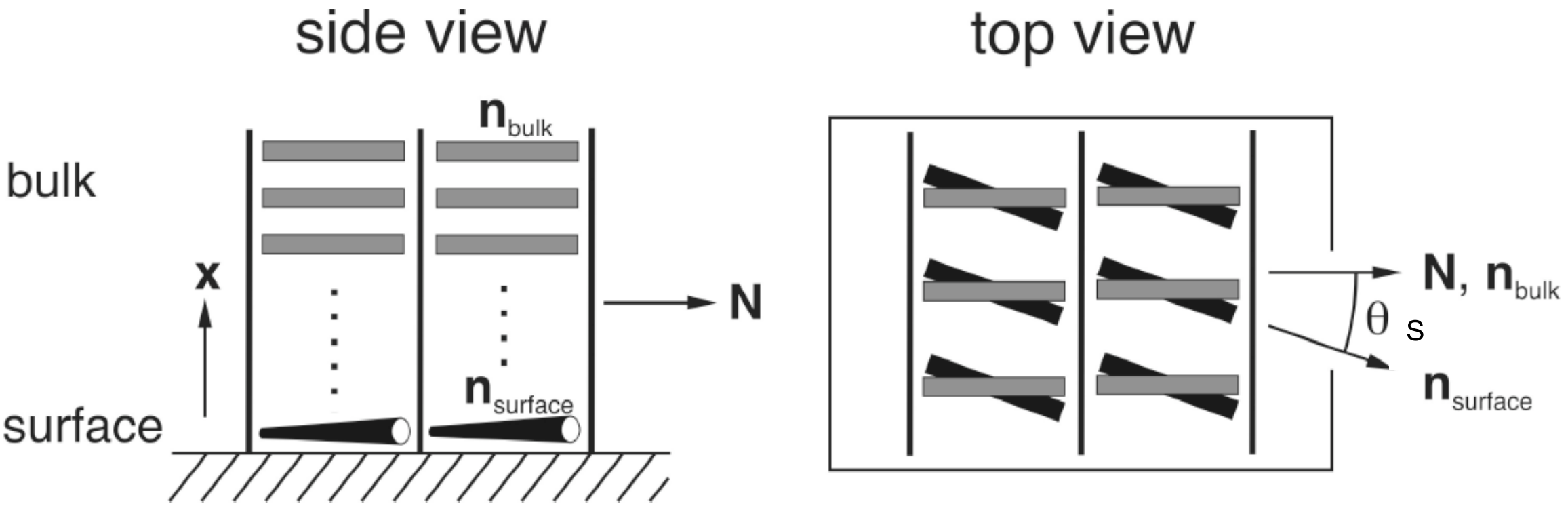}
\caption{A schematic of the surface electroclinic effect (SECE), whereby a coupling between molecular dipoles and a surface induces local tilt of the director ${\bf \hat n}$, away from the smectic layer normal ${\bf \hat N}$ at the surface. Conversely, on rubbed surfaces the SECE makes the bulk smectic normal ${\bf \hat N}$ form at an angle $\theta_S$ to the rubbing direction ${\bf \hat n_{surface}}$.}
\label{SECESchematic}
\end{center}
\end{figure}

The surface electroclinic effect (SECE) is a surface analog of the BECE whereby a coupling between molecular dipoles and a surface induces local tilt of the director ${\bf \hat n}$ away from the smectic layer normal ${\bf \hat N}$ at the surface, as shown in Fig.~\ref{SECESchematic}. The SECE has been analyzed extensively, both experimentally and theoretically, for materials with a continuous Sm-$A^*$ -- Sm-$C^*$ transition \cite{Xue&Clark,Chen&Ouchi,Rovsek&Zeks, Shao&Boulder}. In sample cells in which one of the two polymer coated glass plates is rubbed to align the director ${\bf \hat n}$ at the surface (as shown in Fig.~\ref{SECESchematic}) the SECE makes the smectic layer normal  ${\bf \hat N}$ deviate by an angle $\theta_S$ from the rubbing direction, i.e., from ${\bf \hat n_{surface}}$. This angle $\theta_S$ can be measured using polarized light microscopy.  However, surface pinning prevents the once-formed layer structure from rotating which means that the surface tilt angle $\theta_S$ is effectively stuck at the temperature, $T_A$, where the layers first form in the Sm-$A^*$ phase \cite{Chen&Ouchi}. Thus for such sample cells one cannot easily explore the variation of the SECE with $T$. We note that $T_A$ could perhaps be varied by quenching the system into the Sm-$A^*$ phase at lower temperatures \cite{SECE with Varying T}. 

The BECE can be explored by varying $E$ and $T$, and for the SECE the analogous parameters would be surface coupling $w$ (to be defined explicitly later) and $T$, although as discussed above, varying $T$ is nontrivial. Surface coupling could presumably be varied through differing surface treatments. As we will see in Section \ref{Model and Analysis}, the surface coupling $w$ is an odd, monotonically increasing function of enantiomeric excess (i.e., chirality). Thus $w$ could also be varied via the enantiomeric excess. However, one must be careful because we will see that variation of enantiomeric excess will also change the Sm-$A^*$ -- Sm-$C^*$ transition temperature. Correspondingly, at fixed temperature, increasing enantiomeric excess will lead to an increase in $w$ {\it and} a decrease in reduced temperature $\tau$.
\begin{figure}
\includegraphics[scale=1]{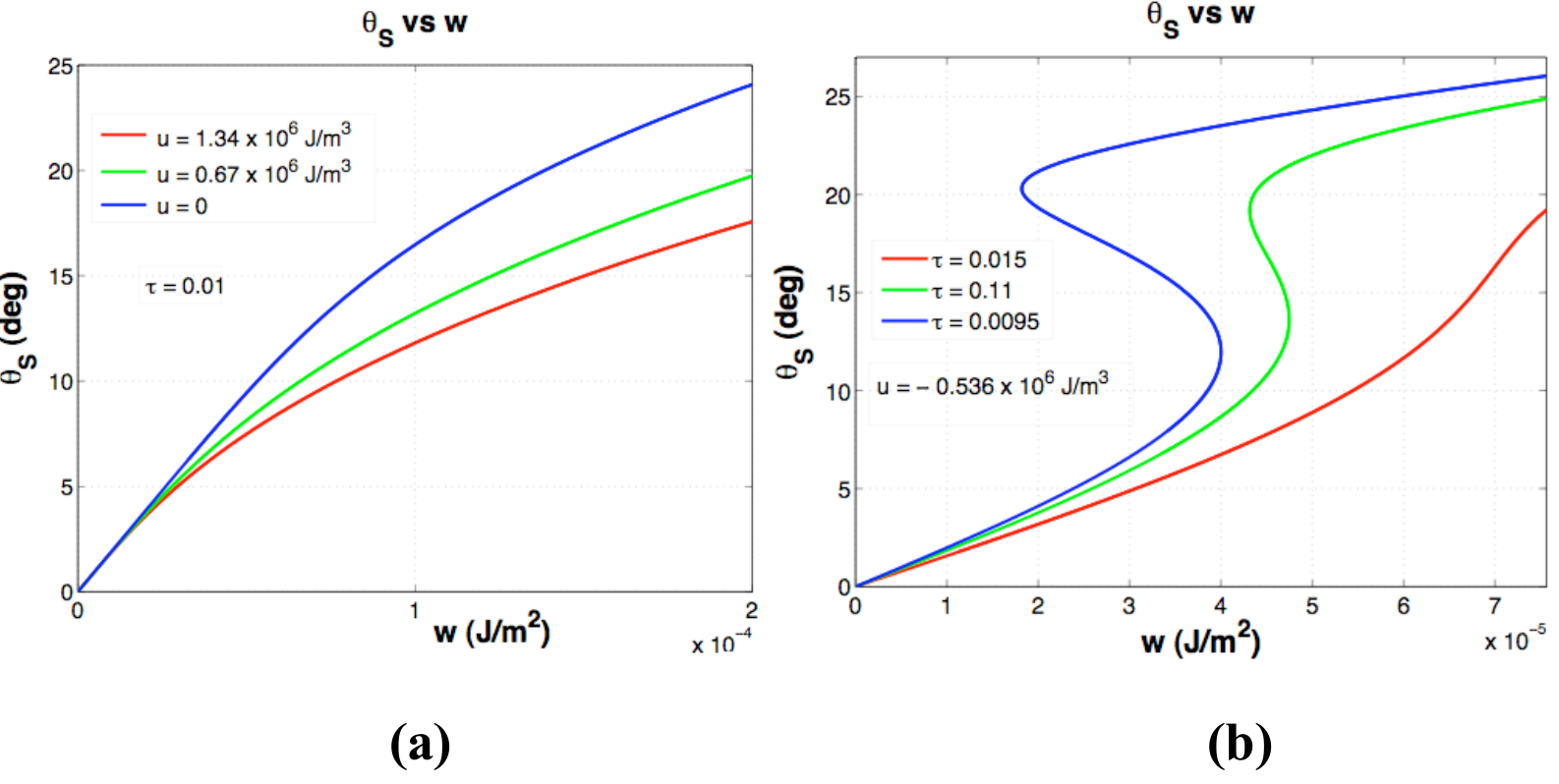}
\caption{Different types of surface electroclinic response curves $\theta_S(w)$, where $w$ is the surface coupling which acts like a field localized at the surface. These are obtained using the theoretical model described in Section \ref{Bulk Electroclinic Effect and Surface Electroclinic Effect}. (a) $\theta_S(w)$ curves for materials in the Sm-$A^*$ phase at $1\%$ reduced temperature (i.e., $\tau = (T-T_{AC})/T_{AC})=0.01$) above a continuous Sm-$A^*$ -- Sm-$C^*$ transition. Each of the three curves corresponds to a material with a transition at a different proximity to a tricritical point, (TCP) with upper/lower (blue/red) being the closest to/furthest from the TCP, respectively. The parameter $u$ is a quantitative measure of this proximity and is defined via Eq.~(\ref{f_u}). One can see that the SECE is significantly enhanced with proximity to the TCP. (b) $\theta_S(w)$ for a material in the Sm-$A^*$ phase above a first order Sm-A$^*$ -- Sm-C$^*$ transition, i.e. with $u<0$. The lower (red) curve corresponds to $T>T_{cs}$ (i.e $\tau>\tau_{cs}$), with $T_{cs}$ a critical temperature, larger than the analogous critical temperature for the BECE. In this case the curve is continuous but ``superlinear'', corresponding to positive curvature at small fields followed by negative curvature at large fields. The upper (blue) curve shows the surface tilt for $T<T_{cs}$ (i.e $\tau<\tau_{cs}$). Now the response is discontinuous, thus predicting a jump as the surface coupling $w$ is varied.}
\label{SECE_Response_curves}
\end{figure}

Recently, we developed a generalized model, which can be used to analyze the SECE near {\it both} continuous and first order Sm-$A^*$ -- Sm-$C^*$ transitions \cite{Rudquist&Saunders}. This was motivated in part by the increasing number of compounds exhibiting a first order transition and also by the dramatic nature of the BECE near such a transition. The predictions of the model are summarized in Fig.~\ref{SECE_Response_curves}. These response curves are analogous to those of Fig.~\ref{BECE_Response_curves} for the BECE but with surface coupling $w$ playing the role of electric field $E$. In the Sm-$A^*$ phase, near a continuous Sm-$A^*$ -- Sm-$C^*$ transition, the induced surface tilt $\theta_S$ is a continuous function of surface coupling (and is thus also a continuous function of enantiomeric excess) and it increases with proximity to a tricritical point. In the Sm-$A^*$ phase, near a first order Sm-$A^*$ -- Sm-$C^*$ transition, the curves are more dramatic. For larger temperatures the $\theta_S(w)$ curves will be superlinear, and at lower temperatures $\theta_S(w)$ will exhibit a jump as the surface coupling $w$ is varied. The $\theta_S(w)$ curves in Fig.~\ref{BECE_Response_curves}(a) are in good qualitative agreement with the experimentally obtained variation of $\theta_S$ with enantiomeric excess for the compound W415 \cite{Shao&Boulder} which exhibits a continuous Sm-$A^*$ -- Sm-$C^*$ transition \cite{415Footnote}. At this time, the SECE near a first order Sm-$A^*$ -- Sm-$C^*$ transition has not yet been studied experimentally, but we hope to see such a study carried out in the future.

Several years ago it was demonstrated experimentally \cite{MaclennanFieldControl} that the surface tilt can be controlled by an electric field, i.e., via the bulk electroclinic effect (BECE). It was shown that the surface tilt can in fact be eliminated and even reversed. As mentioned above, the material studied (W415) has a continuous Sm-$A^*$--Sm-$C^*$ transition, and a corresponding theoretical model for the SECE-BECE combination near continuous transitions was presented. Possible applications, including electrically controlled phase plates and color filters were also discussed.

In this article we present a generalized model for the SECE-BECE combination that can be applied to systems near a continuous {\it or} a first order Sm-$A^*$--Sm-$C^*$ transition. This model allows us to predict the dependence of the surface tilt on the applied electric field $E$. Our results for continuous Sm-$A^*$--Sm-$C^*$ transitions are in good qualitative agreement with those found experimentally \cite{MaclennanFieldControl}. For a first order Sm-$A^*$ -- Sm-$C^*$ transition, our model makes some interesting predictions, one of which is that the surface tilt can jump as $E$ is varied. This is not so surprising given that for a first order transition the BECE and SECE each individually show jumps as $E$ or surface coupling are varied. However, when {\it both} SECE and BECE are combined, both the surface tilt {\it and} bulk tilt will vary with $E$. Each can exhibit a jump but not at the same value of $E$ which means that there will be significant discontinuities in the difference between surface and bulk tilt as $E$ is varied. With regard to technological, e.g., display, applications, this could be a feature to be avoided or potentially exploited.

As discussed above, the surface tilt is effectively stuck at the value it has upon formation of the layers, i.e., upon entry to the Sm-$A^*$ phase, and is not expected to vary much with temperature $T$. Nonetheless, the {\it reduced} temperature $\tau$ of the system may be varied by changing the enantiomeric excess of the system. Increasing the enantiomeric excess will raise the Sm-$A^*$--Sm-$C^*$ transition temperature, which will decrease $\tau$. One must be careful however to account for a corresponding increase in surface coupling $w$.  Using our model we are able to analyze the effect of varying enantiomeric excess and show that it can significantly affect the field dependence of the surface tilt, in particular near a first order Sm-$A^*$--Sm-$C^*$ transition. In this case increasing enantiomeric excess could change the field dependence of surface tilt from continuous to discontinuous.

Our model also allows us to analyze the layer spacing (which depends on the tilt) throughout the cell, and also how the spacing is affected by applied field $E$. This in turn allows us to estimate the strain resulting from a difference between the layer spacing at the surface and in the bulk. We are able to do so, for both continuous and for first order Sm-$A^*$--Sm-$C^*$ transitions, and show that for certain ranges of $E$ this strain can result in layer buckling which reduces the overall quality of the liquid crystal cell. This could explain the experimental observation by Maclennan et al \cite{MaclennanFieldControl} of a sudden drop in the optical properties of a W415 cell as $E$ is varied. 

Of course, the effect of tilt on layer spacing has been a topic of intense interest in the context of de Vries materials \cite{LemieuxR}. In fact, several de Vries materials show a strong BECE which can be attributed to their Sm-$A^*$--Sm-$C^*$ transition being close to tricriticality or first order. (Such a strong BECE with its fast analog electro-optic characteristics makes such materials technologically promising.) For de Vries materials, with small tilt-induced change in layer spacing, the induced strain for a given surface tilt should be smaller. However, the more de Vries-like the material is, the closer the transition will be to tricriticality or first order. This in turn means that the induced tilt will be bigger. Thus the small layer contraction may be offset by large tilt. 

The article is organized as follows. In Section \ref{Model and Analysis} we present the our model that incorporates both the bulk and surface electroclinic effects. We first briefly consider each individually, and discuss how Figs.~\ref{BECE_Response_curves} and \ref{SECE_Response_curves} are obtained. We then consider the combination of BECE and SECE, and obtain the field dependence of surface tilt ($\theta_S(E)$) for both continuous and first order Sm-$A^*$ -- Sm-$C^*$ transitions. In Section \ref{Effects of Varying Enantiomeric Excess} we investigate how variation of enantiomeric excess affects $\theta_S(E)$, and in Section \ref{Strain Effects and Onset of Layer Buckling} we analyze strain effects and buckling. As part of this analysis we look at de Vries considerations. Lastly, we make some concluding remarks in Section \ref{Summary}.

\section{Model and Analysis}
\label{Model and Analysis}

To analyze the BECE and SECE within a liquid crystal cell, we consider the standard, simplified, geometry (shown in Fig.~\ref{SECESchematic}) with the liquid crystal in contact with a single surface at $x=0$ and extending to $x=\infty$. This is valid if the two cell surfaces are separated by a distance $L_x\gg\xi_x$, where $\xi_x$ is a correlation length to be defined below. With this geometry the layer normal ${\bf \hat N}$ will tilt away from the rubbing direction ${\bf \hat n_{surface}}$ within the $yz$ plane by an amount $\theta(x)$. To analyze the SECE and BECE in the Sm-$A^*$ phase we employ a Landau expansion of the bulk and surface free energies, $F_B$ and $F_S$ respectively:
\begin{eqnarray}
F_B= A_\perp \int_0^{\infty} dx \left[f(\theta) + \frac{1}{2\chi} P^2 - \gamma P \theta -E P \right]\;
\label{Bulk free energy}
\end{eqnarray}
and
\begin{eqnarray}
F_S= -A_\perp  E_s l_s P(x=0)\;,
\label{Surface free energy}
\end{eqnarray}
where $A_\perp$ is the area of the surface and $E$ is the applied electric field which is perpendicular to the surface. It can be positive or negative. $P(x)$ is the component of the average polarization perpendicular to the surface and $\gamma$ is a $\theta$-$P$ coupling constant, which is an odd, monotonically increasing function of enantiomeric excess. The coefficient $\chi$ is a generalized d.c. electric susceptibility. The polar surface anchoring strength is expressed as a product of an effective surface field $E_s$ and the (short) length scale $l_s$ over which it acts. Thus we define $V_s\equiv E_s l_s$ to be an effective surface voltage which presumably depends on the surface treatment. In keeping only terms of order $P^2$ we make the standard assumption that the Sm-$A^*$ - Sm-$C^*$ transition is primarily driven by $\theta$, with $P$ playing a secondary role. 

The piece $f(\theta)$ is given by:  
\begin{eqnarray}
f(\theta)=\frac{1}{2} a(T) \theta^2 + \frac{1}{4} u \theta^4+\frac{1}{6} v \theta^6 + \frac{1}{2} K_x \left({\partial_x\theta}\right)^2\;,
\label{f_u}
\end{eqnarray}
where $a(T)=r(T-T_0)/T_0$ and $v>0$. For $u\geq0$ the racemic, i.e., $\gamma=0$, bulk Sm-$A$ - Sm-$C$ transition is continuous and takes place at $T_{2nd}=T_0$ while for $u<0$ the transition is first order and takes place at $T_{1st}=T_0\left( 1+\frac{3 u^2}{16 v r}\right)$ \cite{racemic first order transition}.  To lowest order $K_x=K_T$, the twist elastic constant, and controls the variation of $\theta(x)$, over a length scale $\xi_x=\sqrt{K_T/a}$ along $x$ into the bulk. We do not include elastic energy contributions due to the spatial variation (along $x$) of the layer spacing. The role played by the smectic layers will be considered in Section \ref{Strain Effects and Onset of Layer Buckling}. 

Setting $P$ equal to its minimum value $P_{\min}=\chi \gamma  \theta + \chi E$ leads to a $\theta$-only dependent free energy $F=F_B+F_S$ with :
\begin{eqnarray}
F_B= A_\perp \int_0^{\infty} dx \left[f_e(\theta) -eE\theta \right]\;
\label{Bulk free energy1}
\end{eqnarray}
and 
\begin{eqnarray}
F_S= -A_\perp   V_s e \theta_S \;,
\label{Surface free energy1}
\end{eqnarray}
where we have dropped the $\theta$ independent terms in $F_B$ and $F_S$, and  $\theta_S\equiv\theta(x=0)$. The quantity $e\equiv \chi \gamma$ is a measure of the coupling between tilt and field. It is also an odd, monotonically increasing function of enantiomeric excess. Thus increasing enantiomeric excess leads to stronger field effects. The piece $f_e(\theta)$ has the same form as $f(\theta)$ in Eq.~(\ref{f_u}) but with an $e$ dependent coefficient $a(T,e)=r\tau(T,e)$. The reduced temperature $\tau(T,e)$ is given by
\begin{eqnarray}
\tau(T,e)=\left(\frac{T}{T_0}-1-\frac{e^2}{\chi r}\right)\;.
\label{tau}
\end{eqnarray}

As discussed in the Introduction, the surface tilt $\theta_S$ is effectively stuck at whatever value it takes on once the layers form, i.e., at the temperature $T_A$ where the system enters the Sm-$A^*$ phase. This means that the value of reduced temperature $\tau(T,e)$ in our model, is effectively fixed at $\tau_A=\tau(T_A, e)$. Given that the continuous Sm-$A^*$--Sm-$C^*$ transition takes place at $\tau=0$, $\tau_A$ is really a dimensionless measure of the temperature range of the Sm-$A^*$ phase. A first order Sm-$A^*$--Sm-$C^*$ transition takes place at $\tau_{1st}=\frac{3 u^2}{16 v r}$, so for materials with such a first order transition the difference, $\tau_A-\tau_{1st}$, corresponds to the reduced temperature width of the Sm-$A^*$ phase. For notational simplicity, we will use $\tau$ throughout this article with the understanding that it is fixed at $\tau_A$.

The $e$ dependence of $\tau$ means that $\tau$ can be tuned by varying enantiomeric excess. From Eq.~(\ref{tau}) one sees that increasing the enantiomeric excess will reduce $\tau$, corresponding to an upward shift in the Sm-$A^*$--Sm-$C^*$ transition temperature. For the remainder of this section we will consider a fixed enantiomeric excess such that $e=0.0032$ J/(C m rad), but will discuss the effects of varying enantiomeric excess in Section \ref{Effects of Varying Enantiomeric Excess}. We also consider fixed, typical, values of  $r=1.78\times10^6$ J/(m$^3$ rad$^2$), $v=3.29\times10^6$ J/(m$^3$ rad$^2$), $K=5 \times 10^{-12}$ N. Depending on the situation we will consider different values of $u$, $\tau$, $E$ and $V_s$. 

\subsection{Bulk Electroclinic Effect}
\label{Bulk Electroclinic Effect}

It is useful to first analyze the BECE alone. To do so we simply set the surface voltage $V_s$ equal to zero. In this case the bulk tilt $\theta_B$ is uniform and it is only necessary to deal with the uniform free energy density:
\begin{eqnarray}
f_U(\theta_B)=\frac{1}{2} r \tau \theta_B^2 + \frac{1}{4} u \theta_B^4+\frac{1}{6} v \theta_B^6 -e E \theta_B  \;,
\label{f_BECE}
\end{eqnarray}
Minimization with respect to $\theta_B$ yields
\begin{eqnarray}
e E=r\tau \theta_B + u \theta_B^3+ v \theta_B^5  \;,
\label{theta_B}
\end{eqnarray}
which can be used to obtain the $\theta_B$ vs $E$ response curves. For $u \geq 0$ (i.e., for a system with a continuous Sm-$A^*$--Sm-$C^*$ transition) the curves are shown in Fig.~\ref{BECE_Response_curves}(a). To produce these curves we chose a relatively small $\tau=0.01$ and a range of $u$ values between zero and $1.34\times10^6$ J/(m$^3$ rad$^2$). For all values of $u \geq 0$ the response $\theta_B$ varies continuously with $E$. As $u$ decreases towards zero (towards tricriticality) the response $\theta_B(E)$ grows. Generally, de Vries smectics, which typically have transitions close to tricriticality, have correspondingly low $u$ values.

For a negative $u=-0.536\times10^6$ J/(m$^3$ rad$^2$) (i.e. for a system with a first order Sm-$A^*$--Sm-$C^*$ transition) the curves are shown in Fig.~\ref{BECE_Response_curves}(b). With this value of $u$ one finds $\tau_{1st}=0.0092$. The $\tau$ values chosen to produce the curves shown in Fig.~\ref{BECE_Response_curves}(b) are all larger than this value of $\tau_{1st}$ and thus, each corresponds to the system being in the Sm-$A^*$ phase. For temperatures sufficiently large, i.e., $\tau>\tau_{cb}\equiv\frac{12}{5}\tau_{1st}$, the response is continuous but exhibits what has been termed ``superlinear growth". As shown in Fig.~\ref{BECE_Response_curves}(b), this corresponds to positive curvature at small fields followed by negative curvature at large fields. It can also be seen that the susceptibility $\frac{d\theta_B}{dE}$ is largest at the field where the curvature changes sign. As $\tau$ is reduced towards $\tau_{cb}$ this susceptibility diverges. For $\tau<\tau_{cb}$ the response becomes discontinuous, as shown in Fig.~\ref{BECE_Response_curves}(b), and there is now switching in $\theta_B$ at $E_j$. The value of $E_j$ is temperature dependent and decreases continuously to zero as $\tau$ is lowered towards $\tau_{1st}$.

\begin{figure}
\begin{center}
\includegraphics[scale=0.6]{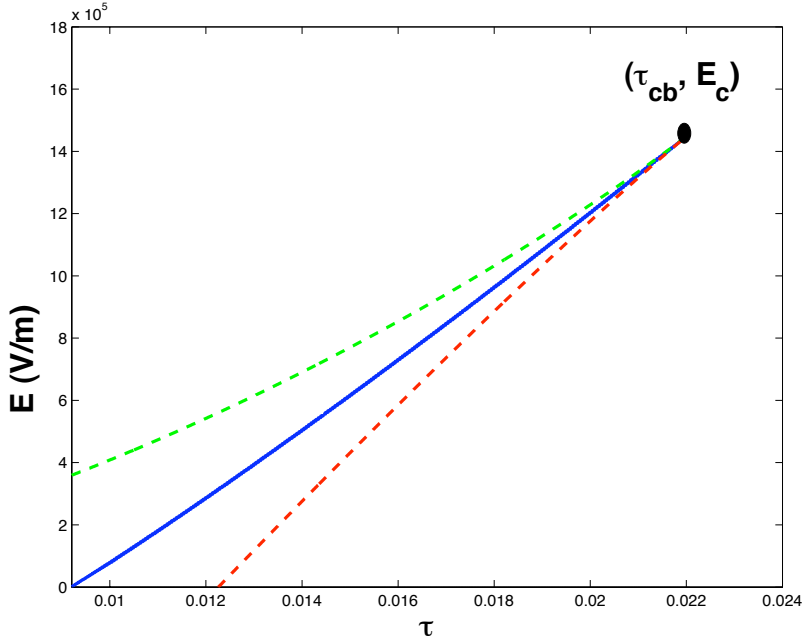}
\caption{The $E$-$\tau$ state map for bulk tilt in a system with a first order Sm-$A^*$--Sm-$C^*$ transition. Low and high tilt states are separated by the middle (blue) line, $E_j(\tau)$, corresponding to a line of first order low tilt -- high tilt Sm-$C^*$ -- Sm-$C^*$ transitions. Low and high tilt states lie below and above the line, respectively. The line extends from ($\tau_{1st}$ , $0$) and terminates at the critical point ($\tau_{cb}=\frac{12}{5}\tau_{1st}$ , $E_c$). The two other lines correspond to the upper (green) and lower (red) metastability boundaries, $E_U(\tau)$ and $E_L(\tau)$ respectively, between which $\theta_B(E)$ is multivalued. See text for further details.}
\label{BulkStateMap}
\end{center}
\end{figure}

To obtain the temperature dependence of $E_j(\tau)$, we recognize that the bulk tilt jumps from (to) the lower $\theta_{BL}$ branch and to (from) the upper $\theta_{BU}$ branch when the free energy of $\theta_{BU}$ becomes less (more) than that of $\theta_{BL}$.  Thus the corresponding value of $E_j(\tau)$ can be found using Eq.~(\ref{theta_B}) and the condition $f_U(\theta_{BL})=f_U(\theta_{BU})$ with $f_U(\theta_B)$ given by Eq.~(\ref{f_BECE}). For the parameter values used to produce Fig.~\ref{BECE_Response_curves}(b) $E_j=0.7 \times 10^5$ V/m. Finding $E_j(\tau)$ allows the construction of a state map in $E$-$\tau$ space which is shown in Fig.~\ref{BulkStateMap}. The $E_j(\tau)$ curve, corresponding to a line of first order low tilt -- high tilt Sm-$C^*$ -- Sm-$C^*$ transitions, terminates at the critical point ($\tau_{cb}$,$E_c$).  Also shown on this state map are the upper and lower metastability boundaries, $E_U(\tau)$ and $E_L(\tau)$ respectively, between which $\theta_B(E)$ is multivalued. For a quasi-static ramping of the field the system will display hysteresis for $\tau<\tau_{cb}$, and the width of the hysteresis loop will be $\Delta_E=E_U(\tau)-E_L(\tau)$.

\subsection{Bulk Electroclinic Effect and Surface Electroclinic Effect}
\label{Bulk Electroclinic Effect and Surface Electroclinic Effect}

Next we consider the combination of bulk and surface electroclinic effects by analyzing the full free energy $F=F_B+F_S$, with $F_B$ and $F_S$ given by Eqs.~(\ref{Bulk free energy1}) and (\ref{Surface free energy1}). The value of tilt in the bulk, i.e. $\theta(x\rightarrow \infty)=\theta_B$ is governed by $F_B$ alone and is given by Eq.~(\ref{theta_B}). For notational simplicity we define surface coupling $w\equiv  V_s e$ so that $F_s=-A_\perp w \theta_S$. However, as discussed below Eq.~(\ref{tau}), we consider fixed enantiomeric excess, and thus fixed $e$, so a varying $w$ is understood to correspond to varying $V_s$ (via surface treatment) and not varying $e$. Throughout this article the surface coupling is taken to be positive, i.e., $w>0$, so that in the absence of an electric field the induced surface tilt $\theta_S>0$. For non-zero electric field  $\theta_S>\theta_B$ and $\theta(x \rightarrow \infty) \rightarrow \theta_{B_+}$, i.e., $\partial_x \theta < 0$. To find the tilt throughout the system we use the Euler-Lagrange equation:
\begin{eqnarray}
\partial_x^2\theta= \frac{1}{K_T}\frac{\partial f_U(\theta)}{\partial \theta}\;,
\label{E-L eqn}
\end{eqnarray}
with $f_U(\theta)$ given by Eq.~(\ref{f_BECE}). The above equation, along with the fact that $\partial_x \theta\rightarrow0$ as $x\rightarrow \infty$, can then be integrated to yield
\begin{eqnarray}
\partial_x \theta=  - \sqrt{\frac{2(f_U(\theta(x))-f_U(\theta_B))}{K_T}}\;.
\label{First integral}
\end{eqnarray}
Using Eq.~(\ref{First integral}) we can express the total free energy  $F=F_B+F_S$ as
\begin{eqnarray}
F= A_\perp\left[\int_{\theta_B}^{\theta_S} \sqrt{2K_T(f_U(\theta)-f_U(\theta_B))}d\theta - w\theta_S\right] \;.
\label{Total free energy ito theta_S}
\end{eqnarray}
Minimizing the above $F$ with respect to $\theta_S$ we find the following implicit equation for the surface tilt $\theta_S$:
\begin{eqnarray}
f_U(\theta_S)=f_U(\theta_B)+\frac{w^2}{2K_T}\;,
\label{theta_S}
\end{eqnarray}
where we note that the electric field $E$ appears in both $f_U(\theta_S)$ and $f_U(\theta_B)$. Together, Eqs.~(\ref{f_BECE}), (\ref{theta_B}) and (\ref{theta_S}) can be used to obtain $\theta_S(E)$. This can be done near either continuous or first order Sm-$A^*$ -- Sm-$C^*$ transitions which we consider separately below.
\begin{figure}
\begin{center}
\includegraphics[scale=0.7]{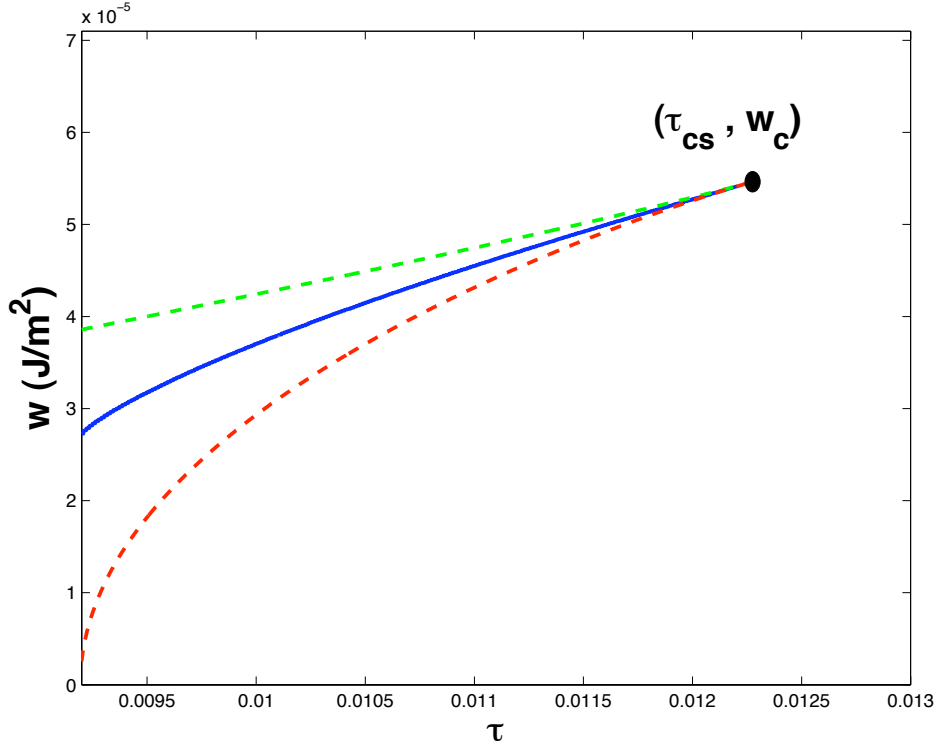}
\caption{The $w$-$\tau$ state map for surface tilt in a system with a first order Sm-$A^*$--Sm-$C^*$ transition. As discussed in the text, it is important to note that reduced temperature $\tau$ is effectively fixed at its value upon entry to the Sm-$A^*$ phase, i.e.,  $\tau=\tau_A=\tau(T_A,\epsilon)$. Thus $\tau_A$ is effectively a measure of the width of the Sm-$A^*$ phase. Low and high surface tilt states are separated by the middle (blue) line $w_j(\tau)$,  corresponding to a line of first order low -- high surface tilt Sm-$C^*$ -- Sm-$C^*$ transitions. Low and high surface tilt states lie below and above the line, respectively. This line terminates at the critical point ($\tau_{cs}$,$w_c$). The two other lines correspond to the upper (green) and lower (red) metastability boundaries, $w_U(\tau)$ and $w_L(\tau)$ respectively, between which $\theta_S(E)$ is multivalued. See text for further details.}
\label{SurfaceStateMap}
\end{center}
\end{figure}

The SECE in zero field is governed by Eq.~(\ref{theta_S}) with $\theta_B=0$, and thus $f_U(\theta_B)=0$. This allows one to generate plots of surface tilt $\theta_S$ vs surface coupling $w$, e.g., Figs.~\ref{SECE_Response_curves}(a) and (b). For a material with a first order Sm-$A^*$ -- Sm-$C^*$ transition one can also obtain a $w$-$\tau$ state map for the surface tilt in zero field, analogous to the bulk tilt state map of Fig.~\ref{BulkStateMap}. This surface tilt state map is shown in Fig.~\ref{SurfaceStateMap}. The threshold surface coupling $w_j(\tau)$, at which the system jumps between low and high surface tilt states, corresponds to the $w$ value at which the free energy of the states are equal. In finding $w_j(\tau)$ one must be careful to compare $F$, the {\it integrated} free energy of Eq.~(\ref{Total free energy ito theta_S}). In finding the threshold field $E_j(\tau)$ for the bulk state map one needed only compare the free energy densities $f$ because the tilt, and thus free energy density, was uniform throughout the system. For a non-zero surface tilt, the tilt and thus the free energy density vary as one moves from surface to bulk.

\subsection{Field Dependence of Surface Tilt Near a Continuous Sm-$A^*$--Sm-$C^*$ Transition}
\label{Field Dependence of Surface Tilt Near a Continuous Sm-$A^*$--Sm-$C^*$ Transition}

\begin{figure}
\begin{center}
\includegraphics[scale=0.7]{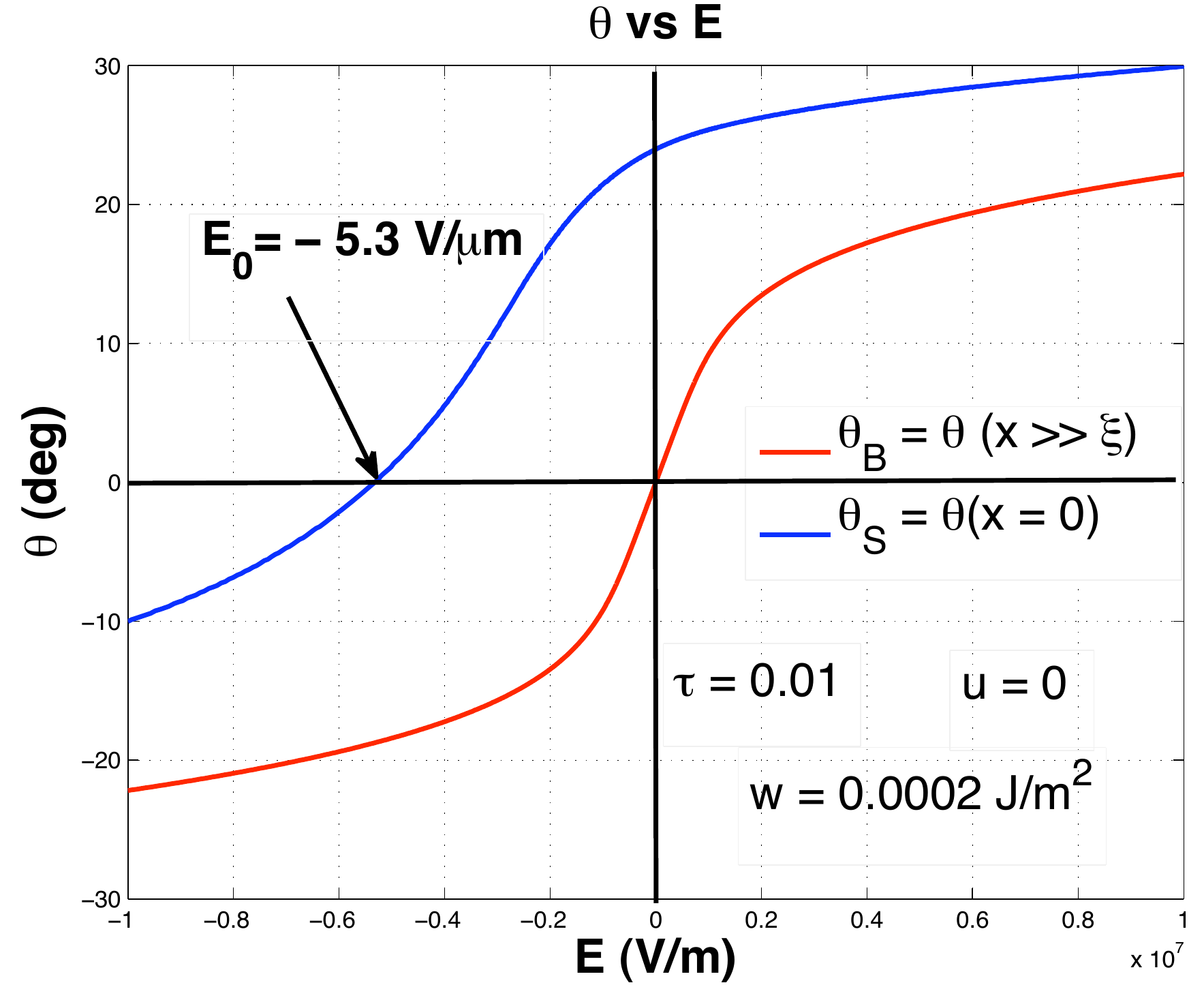}
\caption{The upper (blue) curve shows the surface tilt $\theta_S$ vs electric field $E$ in the Sm-$A^*$ phase near a continuous Sm-$A^*$--Sm-$C^*$ transition. For reference the corresponding bulk tilt $\theta_B$ vs $E$ is also shown as the lower (red) curve. Note that the surface coupling $w$ is positive so that $\theta_S>\theta_B$. The value $E_0$ corresponds to the field that eliminates surface tilt.}
\label{thetaSvsEcontAC}
\end{center}
\end{figure}

Figure ~\ref{BECE_Response_curves}(a) shows that in the Sm-$A^*$ phase, near a continuous Sm-$A^*$--Sm-$C^*$ transition (i.e., for $u>0$), there is only one possible bulk tilt $\theta_B$ for a given value of electric field $E$. The surface tilt will also be a single valued function of $E$. Figure~\ref{thetaSvsEcontAC} shows a sample $\theta_S$ vs $E$ curve, along with the corresponding $\theta_B(E)$ curve, for $\tau=0.01$, $w=2\times10^{-4}$ J/(m$^2$ rad) and $u=0$. That $u=0$ is consistent with the fact that the material W415  (which we want to compare our model with) has a Sm-$A^*$--Sm-$C^*$ transition at, or very near, tricriticality. As $\tau$ is varied the $\theta_S(E)$ and $\theta_B(E)$ curves remain qualitatively similar. 

It can be seen that the surface tilt can be tuned to zero (at $E_0$) or even reversed, consistent with experimental observation \cite{MaclennanFieldControl}. Using the above parameter values and $\tau=0.01$, we find $E_0=-5.3$ V/$\mu$m. Our generalized model also allows us to investigate how $E_0$ varies with reduced temperature $\tau$. As discussed in the Introduction, the surface tilt $\theta_S$ is effectively stuck at whatever value it takes on once the layers form, i.e., at the temperature $T_A$ where the system enters the Sm-$A^*$ phase. This means that the value of reduced temperature $\tau(T,\epsilon)$ in our model is effectively fixed at $\tau_A=\tau(T_A,\epsilon)$ and is really a measure of the temperature range of the Sm-$A^*$ phase. One should, however, keep in mind that $T_A$ could perhaps be varied by quenching the system into the Sm-$A^*$ phase at lower temperatures \cite{SECE with Varying T}. 

\begin{figure}
\begin{center}
\includegraphics[scale=0.9]{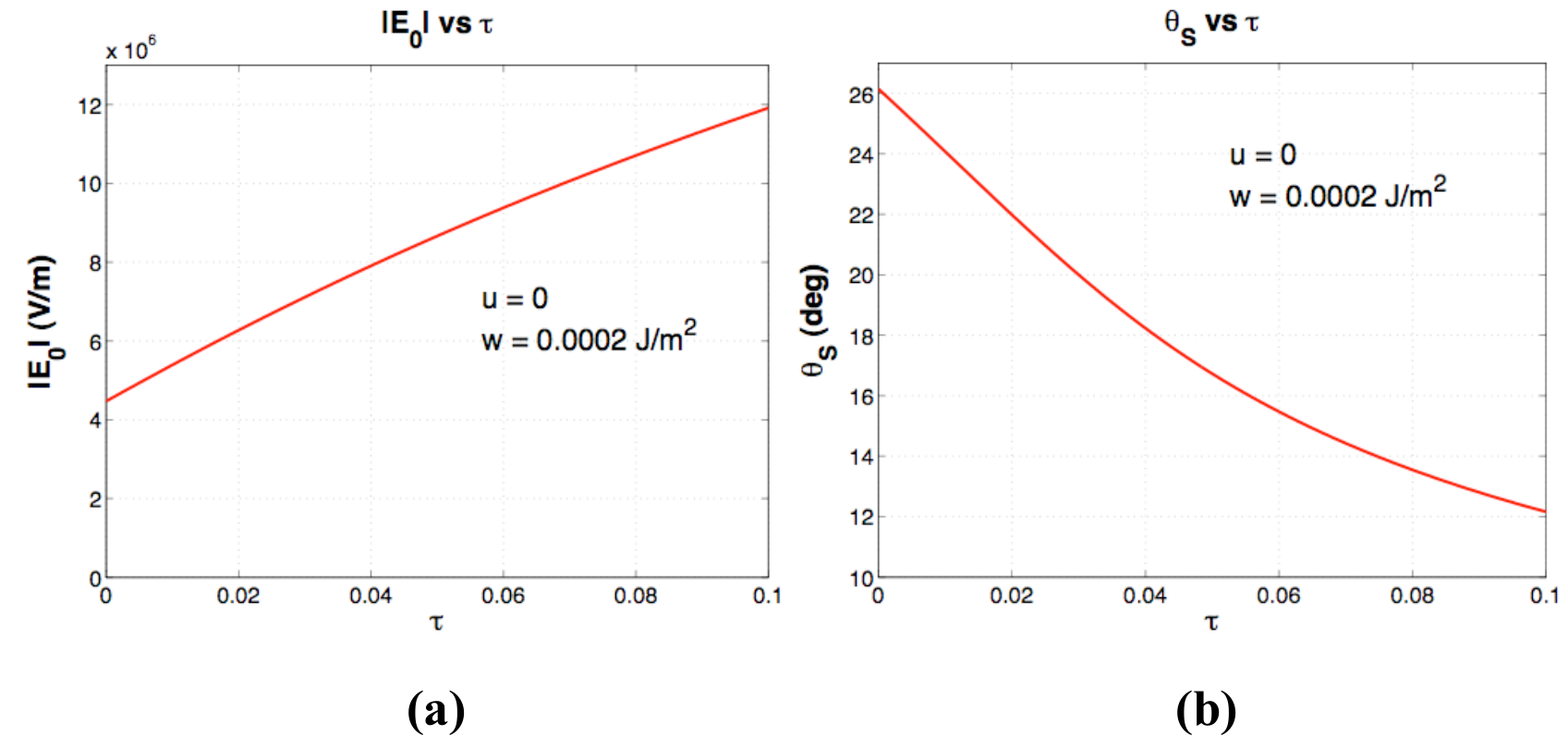}
\caption{Variation with reduced temperature $\tau$ of $|E_0|$, the magnitude of the field required to eliminate the surface tilt. Also shown is the variation with  $\tau$ of $\theta_S(E=0)$ the magnitude of the surface tilt being eliminated. It is interesting to note that even though $\theta_S(E=0)$ grows larger as $\tau$ decreases, the corresponding $|E_0|$ grows smaller. This is discussed further in the text.}
\label{E0_vs_tau_ContAC}
\end{center}
\end{figure}

It is straightforward to obtain $E_0(\tau)$. Using Eq.~(\ref{theta_B}) and Eq.~(\ref{theta_S}) with $\theta_S$ set equal to zero, one can obtain separate equations for $E_0$ and $\tau$, each parameterized in terms of $\theta_B$. Figure~\ref{E0_vs_tau_ContAC} shows the resulting plot of $|E_0|$ vs $\tau$. The parameter values used to produce this plot are the same as those used to produce Fig.~\ref{thetaSvsEcontAC}. For reference we also show a plot of $\theta_S(E=0)$ vs $\tau$, where $\theta_S(E=0)$ is the value of the surface tilt in the absence of an applied field, i.e., the surface tilt that is being eliminated. As expected, $\theta_S(E=0)$ grows with decreasing $\tau$ because the closer one is to the Sm-$C^*$ phase, the larger the surface tilt will be. However, the magnitude of field ($|E_0|$) needed to eliminate this surface tilt {\it decreases} with decreasing $\tau$, i.e., the larger the surface tilt, the {\it smaller} the $|E_0|$ needed to eliminate it. While this may seem counterintuitive it can be understood as follows. While decreasing $\tau$ does increase the surface tilt, it {\it also} increases the bulk electroclinic effect susceptibility, making the application of a bulk field $E$ more effective, thus requiring a smaller eliminating field $|E_0|$. 

\subsection{Field Dependence of Surface Tilt Near a First Order Sm-$A^*$ -- Sm-$C^*$ transition}

For a first order Sm-$A^*$ -- Sm-$C^*$ transition the electric-field dependence of surface tilt will exhibit significant qualitative differences depending on $\tau$. The reader is reminded that $\tau$ in our model, is effectively fixed at $\tau_A=\tau(T_A,\epsilon)$. Since a first order Sm-$A^*$--Sm-$C^*$ transition takes place at $\tau_{1st}=\frac{3 u^2}{16 v r}$, the difference $\tau-\tau_{1st}$, corresponds to the reduced temperature width of the Sm-$A^*$ phase. As discussed in Section \ref{Bulk Electroclinic Effect}, and shown in Fig.~\ref{BECE_Response_curves}, the bulk tilt response $\theta_B(E)$ becomes discontinuous for $\tau<\tau_{cb}\equiv\frac{12}{5}\tau_{1st}$. However, the analogous $\theta_S(w)$ (surface tilt vs surface coupling) curve becomes discontinuous for $\tau<\tau_{cs}\equiv\frac{4}{3}\tau_{1st}$. Since $\tau_{cs}=\frac{5}{9}\tau_{cb}$, the $\theta_S(w)$ curve becomes discontinuous at a lower temperature than for the $\theta_B(E)$ curve. Thus we consider three distinct ranges of $\tau$: Range 1: $\tau>\tau_{cb}$, Range 2: $\tau_{cs}<\tau<\tau_{cb}$, and Range 3: $\tau<\tau_{cs}$. For reasons discussed above, accessing each range of $\tau$ would require a different $\tau_A$, which could be achieved using different chemical homologs. In Section \ref{Effects of Varying Enantiomeric Excess} we will discuss how $\tau_A$ could also be tuned by varying enantiomeric excess. One must be careful in this case because varying enantiomeric excess will also change $w$. For each curve we set $u=-0.536\times10^6$ J/(m$^3$ rad$^2$) and $w=2.5\times10^{-5}$ J/(m$^2$ rad).

\begin{figure}
\begin{center}
\includegraphics[scale=0.75]{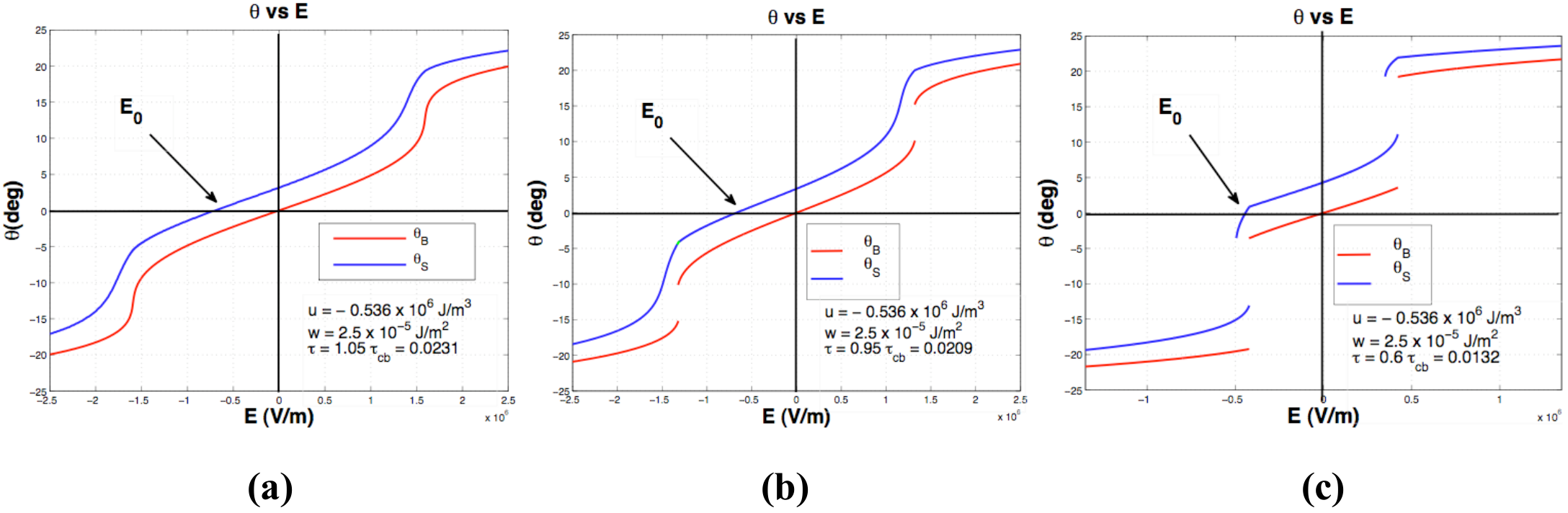}
\caption{Sample surface tilt $\theta_S$ vs electric field $E$ curves in the Sm-$A^*$ phase near a first order Sm-$A^*$--Sm-$C^*$ transition. For reference the corresponding bulk tilt $\theta_B$ vs $E$ is also shown as the lower (red) curve. The three sets of curves correspond to different temperatures, one from each of Ranges 1, 2, 3 discussed in the text. In each case, the surface coupling is positive so that $\theta_S>\theta_B$, and the value $E_0$ corresponds to the field that eliminates surface tilt. (a) Range 1: $\tau=0.0362=1.05\tau_{cb}$. In this case both $\theta_S(E)$ and $\theta_B(E)$ are continuous. (b) Range 2: $\tau=0.95\tau_{cb}$. In this case both $\theta_S(E)$ is continuous but $\theta_B(E)$ is discontinuous and exhibits a jump. (c) Range 3: $\tau=0.6 \tau_{cb}$. In this case both $\theta_S(E)$ and $\theta_B(E)$ are discontinuous. See text for further discussion of these plots.}
\label{thetaSvsE1stAC}
\end{center}
\end{figure}

To find the field dependence of surface tilt, $\theta_S(E)$, in Range 1 we use Eqs.~(\ref{f_BECE}), (\ref{theta_B}) and (\ref{theta_S}). Figure~\ref{thetaSvsE1stAC}(a) shows a sample $\theta_S$ vs $E$ curve, along with the corresponding $\theta_B(E)$ curve. The reduced temperature is $\tau=0.0362$ which corresponds to $\tau=1.05\tau_{cb}$. In this case both $\theta_S(E)$ and $\theta_B(E)$ are continuous.

In Range 2 the bulk tilt response function $\theta_B(E)$ is multivalued and in using Eqs.~(\ref{f_BECE}), (\ref{theta_B}) and (\ref{theta_S}) one must be careful to use the lowest energy $\theta_B$ value, i.e., the $\theta_B$ value yielding the lowest value of $f_U(\theta_B)$. A sample $\theta_S(E)$ curve along with the corresponding $\theta_B(E)$ curve is shown in Fig.~\ref{thetaSvsE1stAC}(b), corresponding to $\tau=0.95\tau_{cb}$. Note that the $\theta_S(E)$ curve is continuous but exhibits a kink at the field values where $\theta_B$ jumps. 

In Range 3 one must also be careful to use the lowest energy $\theta_B(E)$ value when finding $\theta_S(E)$. Figure~\ref{thetaSvsE1stAC}(c) shows a sample $\theta_S$ vs $E$ curve along with the corresponding $\theta_B(E)$ curve, for $\tau=0.6\tau_{cb}$. Now both $\theta_S(E)$ and $\theta_B(E)$ are discontinuous functions of $E$, with each exhibiting a jump as $E$ is varied. As discussed in Section \ref{Bulk Electroclinic Effect}, the value of $E_{jB}$ at which $\theta_B$ jumps from low bulk tilt ($\theta_{BL}$) to high bulk tilt ($\theta_{BU}$) is found using Eq.~(\ref{theta_B}) and $f_U(\theta_{BL})=f_U(\theta_{BU})$, with the free energy density $f_U(\theta_B)$ given by Eq.~(\ref{f_BECE}). From Fig.~\ref{thetaSvsE1stAC}(c) it can be seen that the surface tilt is multivalued for a range of positive $E$ and a range of negative $E$.  The value of $E_{jS}$ at which $\theta_S$ jumps will be somewhere within the multivalued range. Inspecting the figure it can be seen that for the positive field values, $E_{jS} <  E_{jB}$, and that for the negative field values, $|E_{jS}| >  |E_{jB}|$. To find the positive or negative value $E_{jS}$ at which $\theta_S$ jumps one must compare the total, integrated free energy $F$, given by Eq.~(\ref{Total free energy ito theta_S}), for low and high surface tilts. The jump in surface tilt will occur at $E=E_{jS}$ where $F(\theta_{SL})=F(\theta_{SU})$. 

\begin{figure}
\begin{center}
\includegraphics[scale=0.8]{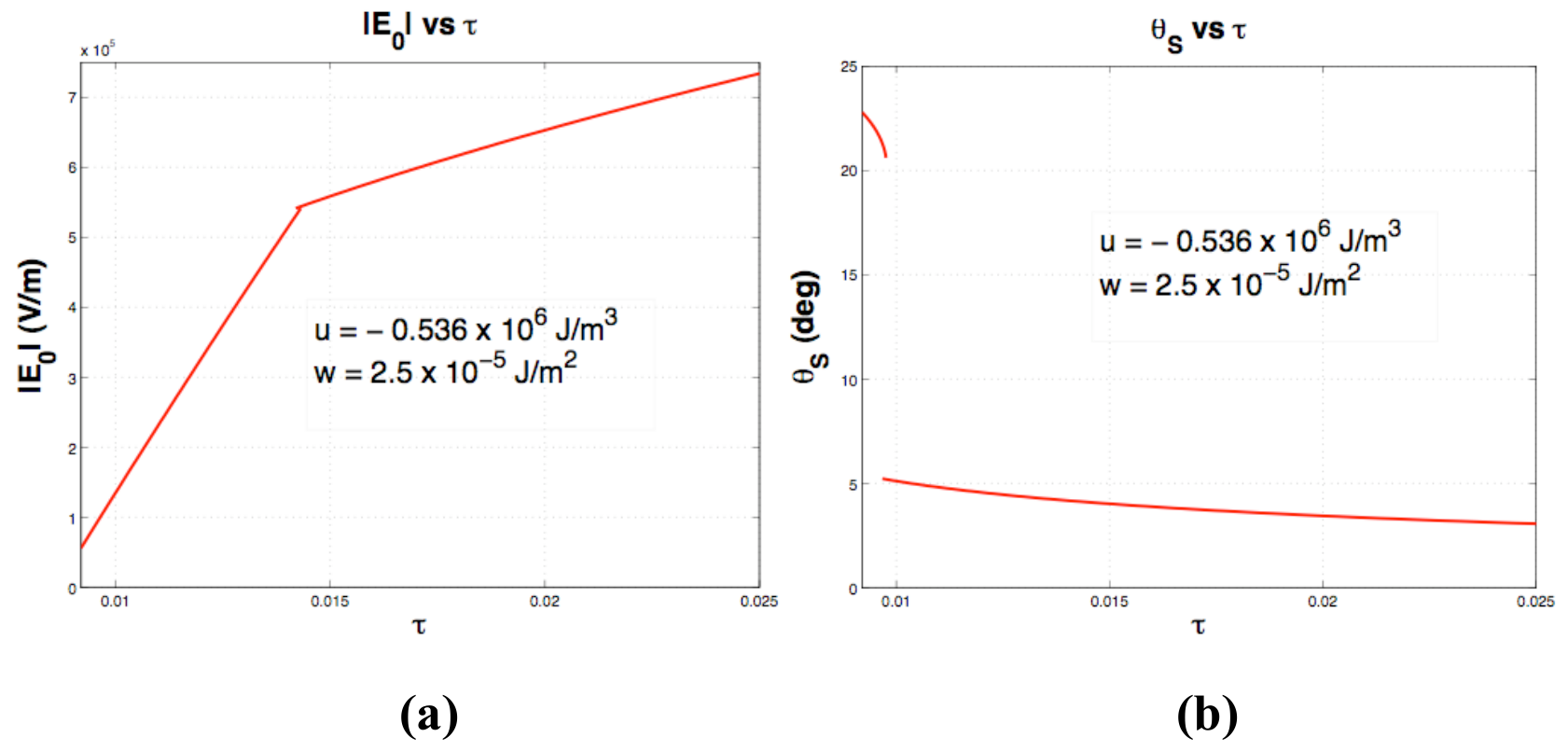}
\caption{Variation with reduced temperature $\tau$ of $|E_0|$, the magnitude of the field required to eliminate the surface tilt. Also shown, is the variation with  $\tau$ of $\theta_S(E=0)$ the magnitude of the surface tilt being eliminated. The kink in the $|E_0|(\tau)$ curve occurs when the magnitude of the bulk tilt $|\theta_B(E=E_0)|$ jumps from low to high. There is a jump in the surface tilt once $\tau\lesssim\tau_{cs}=0.0122$}
\label{E0_vs_tau_1stAC}
\end{center}
\end{figure}

As in Section \ref{Field Dependence of Surface Tilt Near a Continuous Sm-$A^*$--Sm-$C^*$ Transition}, we can analyze how $E_0$ varies with reduced temperature $\tau$. Figure~\ref{E0_vs_tau_1stAC} shows the resulting plot of $|E_0|$ vs $\tau$ along with $\theta_S(E=0)$ vs $\tau$, where $\theta_S(E=0)$ is the value of the surface tilt that is being eliminated. Like the corresponding $|E_0|(\tau)$ curve near a continuous transition, even though $\theta_S(E=0)$ grows larger as $\tau$ decreases, the corresponding $|E_0|$ grows smaller. The reason is the same, and is discussed in Section \ref{Field Dependence of Surface Tilt Near a Continuous Sm-$A^*$--Sm-$C^*$ Transition}. The kink in the $|E_0|(\tau)$ curve occurs when the magnitude of the bulk tilt $|\theta_B(E=E_0)|$ jumps from low to high.

\section{Effects of Varying Enantiomeric Excess} 
\label{Effects of Varying Enantiomeric Excess} 

In this section we discuss the effects of varying the enantiomeric excess which in our analysis so far we have considered fixed. The enantiomeric excess could be varied by mixing enantiomers of opposite handedness, and will reach saturation when the system is made up entirely of a enantiomer with a single handedness. It will be zero for a racemic mixture. 

In our original model, discussed in Section \ref{Model and Analysis} and represented by Eqs.~(\ref{Bulk free energy1}) - (\ref{tau}), the parameters that vary with enantiomeric excess are $e$ and $\tau$. The parameter $e$ represents the coupling between tilt and field: for the bulk tilt it appears in the free energy density as $eE\theta_B$ while for the surface tilt it appears as $w \theta_S$, where $w\equiv  V_s e$.  The parameter $e$ is an odd, monotonically increasing, function of enantiomeric excess. Thus increasing $e$ amplifies the effect of both $E$ and $V_s$. As seen from Eq.~(\ref{tau}), the enantiomeric excess also affects $\tau$ through its dependence on $e$. Increasing the enantiomeric excess (of either handedness) will reduce $\tau(e)$, corresponding to an upward shift in the Sm-$A^*$--Sm-$C^*$ transition temperature. As discussed above, the surface tilt is effectively stuck at whatever value it takes on once the layers form, i.e., at the temperature $T_A$ where the system enters the Sm-$A^*$ phase. Thus $\tau$ cannot be changed through varying temperature $T$. However, the dependence of $\tau$ on $e$ provides another way to vary $\tau$. If one wishes to explore behavior for different $\tau$ values one can simply change the enantiomeric excess. Of course, varying $e$ will also affect surface coupling $w$ and the influence of the bulk field $E$. 
\begin{figure}
\begin{center}
\includegraphics[scale=0.8]{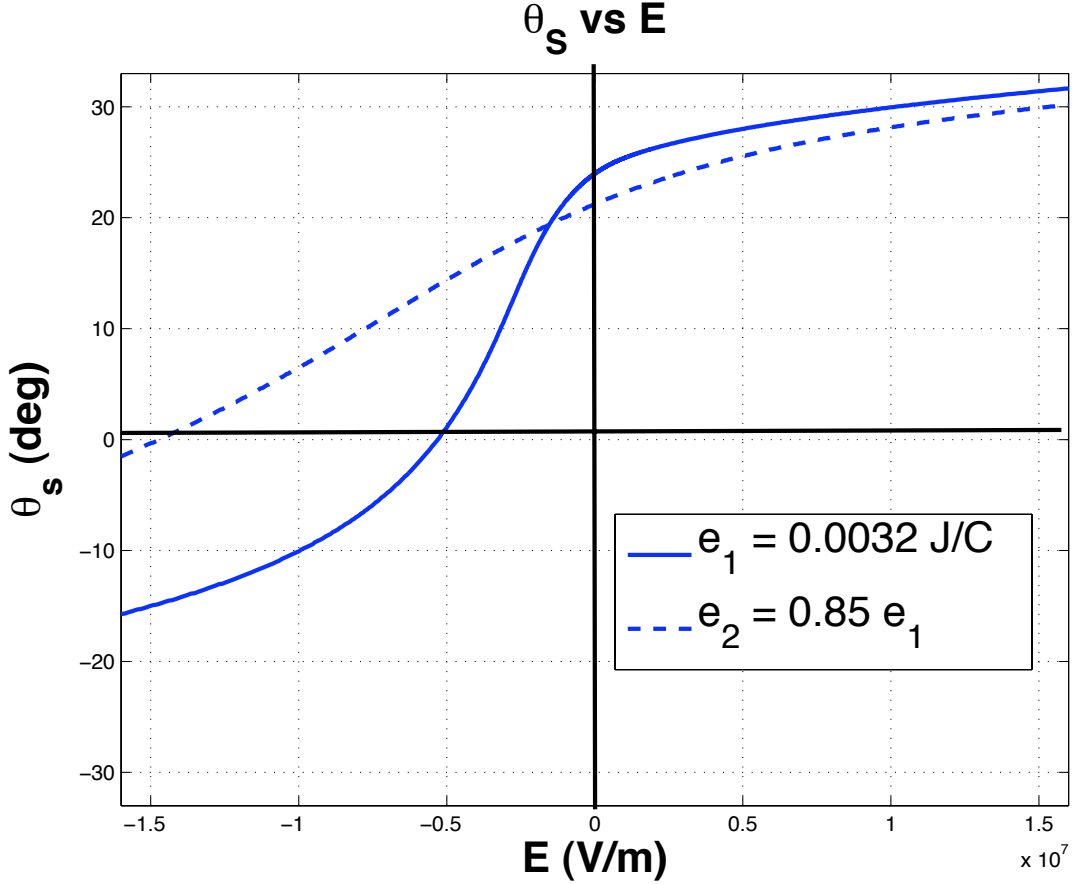}
\caption{Surface tilt $\theta_S$ vs electric field $E$ in the Sm-$A^*$ phase near a continuous Sm-$A^*$--Sm-$C^*$ transition. The dashed curve corresponds to a system with 15$\%$ smaller enantiomeric excess, and the zero field surface tilt is correspondingly smaller. However, the field required to eliminate this smaller surface tilt is larger. This is discussed further in the text. }
\label{u>0ThetaSForDifferentEEs}
\end{center}
\end{figure}

Figure~\ref{u>0ThetaSForDifferentEEs} shows $\theta_S$ vs $E$ near a continuous Sm-$A^*$--Sm-$C^*$ transition for two systems with different enantiomeric excesses, i.e., different $e$ values, $e_1=0.032$ J/C and $e_2=0.85 e_1$. The solid curve is the same as that in Fig.~\ref{thetaSvsEcontAC}. The dashed curve, corresponding to the system with 15$\%$ smaller enantiomeric excess \cite{e and EE}, has a smaller zero-field surface tilt. However, the magnitude of the field required to eliminate this surface tilt is significantly larger. This can be understood qualitatively using Fig.~\ref{E0_vs_tau_ContAC}. The reduction in $e$ causes both a reduction in $w$ (causing surface tilt to decrease) and the effect of $E$ (causing an increase in the field needed to eliminate the surface tilt). These effects essentially offset each other but the decrease in $e$ also leads to an increase in $\tau$. From Fig.~\ref{E0_vs_tau_ContAC} it can be seen that the increase in $\tau$ leads to an increase in the field $|E_0|$ needed to eliminate the surface tilt.
\begin{figure}
\begin{center}
\includegraphics[scale=0.7]{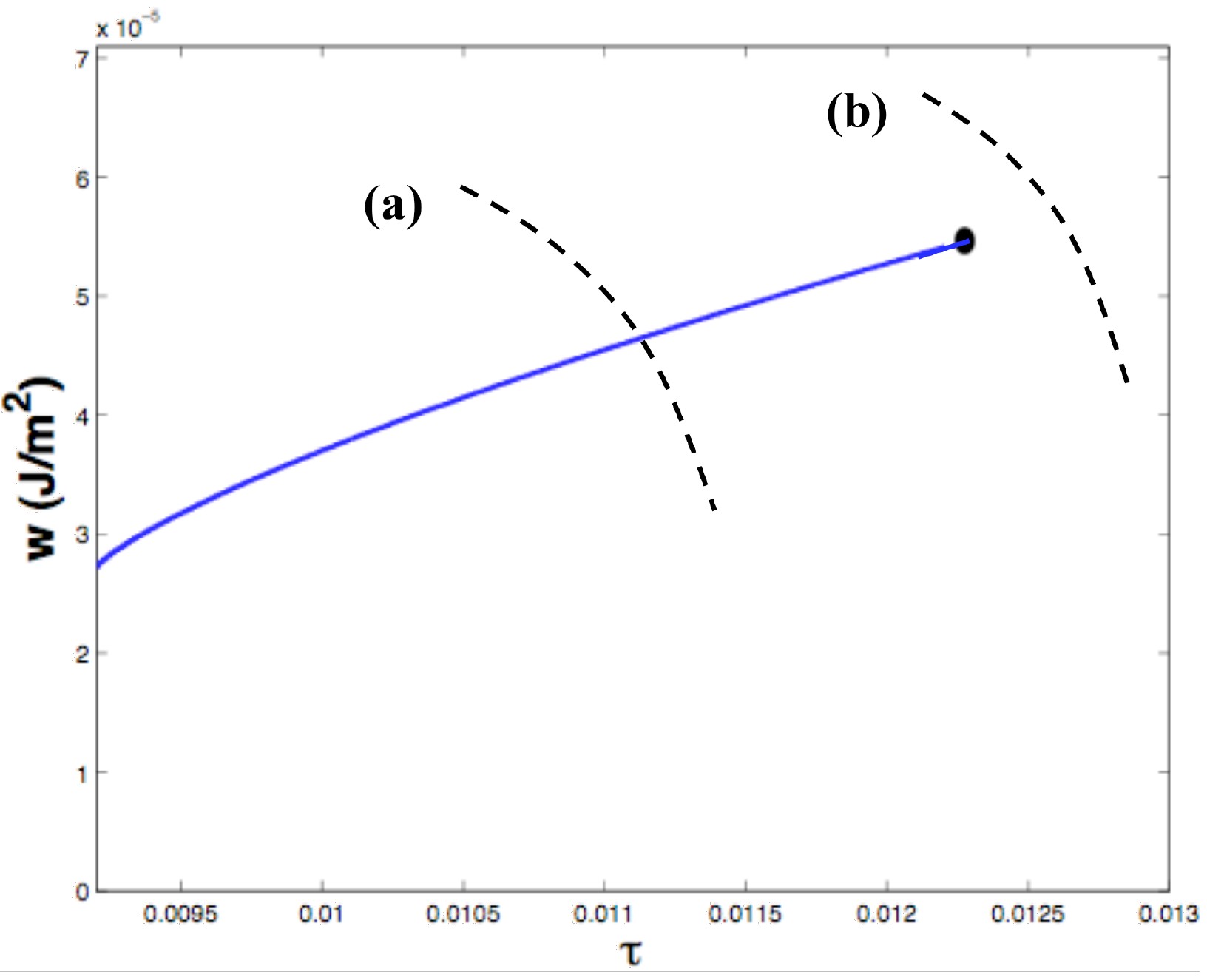}
\caption{The $\tau$-$w$ state map for surface tilt in a system with a first order Sm-$A^*$--Sm-$C^*$ transition in zero applied field. The solid line corresponds to $w_j(\tau)$, the surface coupling at which the surface tilt jumps between high and low states, located above and below the line respectively. Increasing enantiomeric excess corresponds to moving along each locus (dashed lines) from lower right to upper left. Moving along locus (a) will result in the system jumping between low and high surface tilt states. Along locus (b), which lies beyond the critical point, there will be no discontinuity in the surface tilt as enantiomeric excess is varied.}
\label{SurfaceStateMapWithVaryingEE}
\end{center}
\end{figure}

Near a first order Sm-$A^*$--Sm-$C^*$ transition the effect of varying enantiomeric excess can be more dramatic. First consider the case of $E=0$, using the surface tilt state map shown in Fig.~\ref{SurfaceStateMapWithVaryingEE}. For a given surface treatment (i.e., a fixed $V_s$) increasing $e$ will in turn increase $w$ and decrease $\tau$. Thus increasing enantiomeric excess corresponds to moving from lower right to upper left along the locus shown in Fig.~\ref{SurfaceStateMapWithVaryingEE}. Using $\tau$ given by Eq.~(\ref{tau}) and $w\propto e$, it is straightforward to show that the slope magnitude of the locus is proportional to $1/e$. If one starts with a system that at $E=0$ is in a low surface tilt state, one can increase the enantiomeric excess until it crosses the threshold line $w_j$ to a system that is in a high surface tilt state. However, for some systems, the locus of varying enantiomeric excess may lie beyond the critical point in which case there will be no discontinuity between low and high surface tilt states.

\begin{figure}
\begin{center}
\includegraphics[scale=0.8]{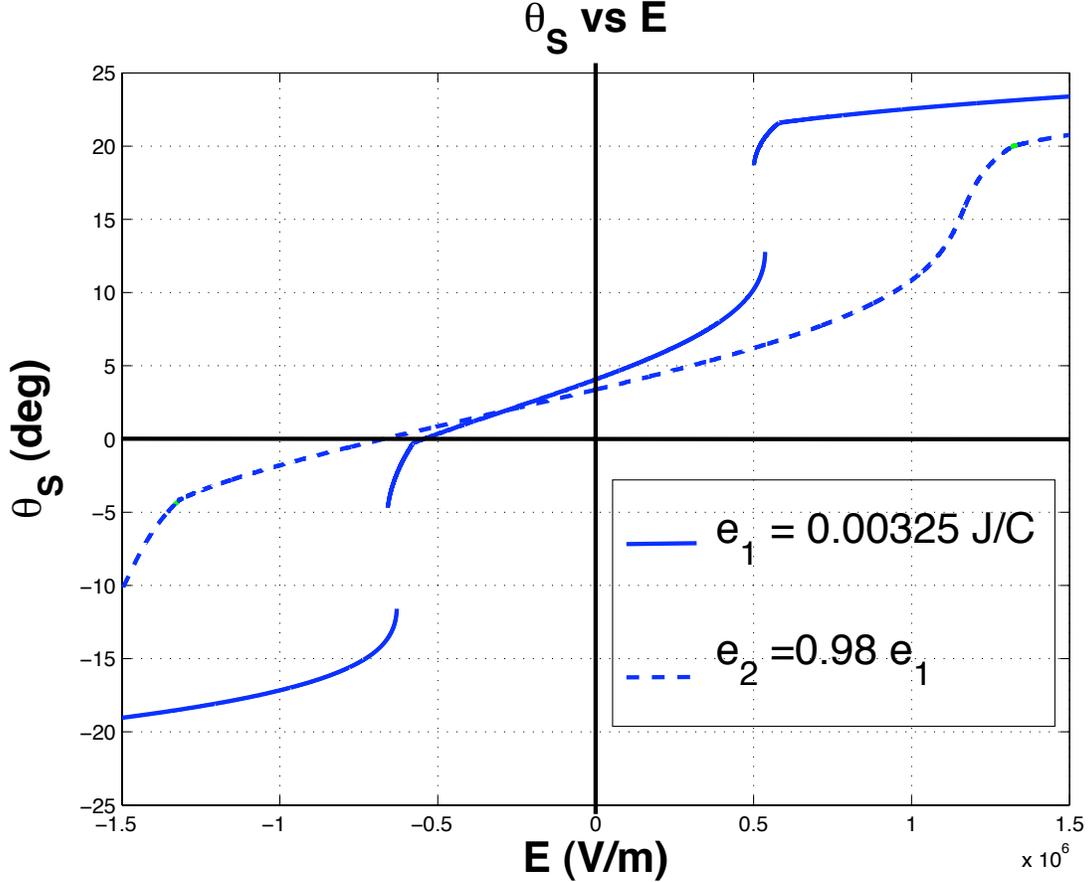}
\caption{Surface tilt $\theta_S$ vs electric field $E$ in the Sm-$A^*$ phase near a first order Sm-$A^*$--Sm-$C^*$ transition. The dashed curve corresponds to a system with 2$\%$ smaller enantiomeric excess, and the zero field surface tilt is correspondingly smaller. The reduction in $e$ significantly changes the nature of the $\theta_S$'s dependence on $E$, from discontinuous to continuous.}
\label{u<0ThetaSForDifferentEEs}
\end{center}
\end{figure}

For nonzero field the effect of varying $e$ is shown in Fig.~\ref{u<0ThetaSForDifferentEEs}. As for the system near a continuous Sm-$A^*$--Sm-$C^*$ transition, a reduction in $e$ leads to a decrease in the zero-field surface tilt and an increase in the field needed to eliminate it. However, the reduction in $e$ significantly changes the nature of the $\theta_S$'s dependence on $E$, from discontinuous to continuous.

\section{Strain Effects and Onset of Layer Buckling}
\label{Strain Effects and Onset of Layer Buckling}

So far we have not discussed the effect that the surface and bulk tilt have on the layers. We do so now and show that for a certain range of applied field $E$ layer buckling may occur, which may decrease the quality of the cell. Before discussing the implications of strain for our system it is useful to consider a simple cell with a non-chiral Sm-$A$ phase in bookshelf geometry. At the top and bottom of the cell are movable plates. Before the plates are moved the layer spacing is at its preferred value $d_p$. If the plates are moved closer together the layers are compressed and are forced to have a layer spacing that is smaller than the preferred spacing $d_p$. To relieve this compressive strain layers must be removed via the formation of dislocations.

On the other hand, if the plates are moved further apart the layers are dilated and are forced to have a layer spacing that is larger than the preferred spacing $d_p$. As with compressive strain, the dilative strain can be relieved through the formation of dislocations, this time to add new layers. However, unlike compressive strain, dilative strain can also be relieved by layer buckling \cite{ClarkMeyer}. This is because buckling, i.e., undulation, of the layers leads to a larger average layer spacing which also relieves the strain. The threshold strain, above which buckling occurs is generally lower than that for dislocation formation. Thus dilative strain is usually relieved by buckling while compressive strain can only be relieved by the removal of layers.

In general the magnitude of the strain $\alpha$ due to an imposed layer spacing $d$ that is different than the preferred value $d_p$ is: 
\begin{eqnarray}
\alpha = 1 - \frac{d_p}{d} \;.
\label{General Strain}
\end{eqnarray}
Thus compressive strain, with $d<d_p$, corresponds to $\alpha<0$, and dilative strain, with $d>d_p$, corresponds to $\alpha>0$. For dilative strain it can be shown \cite{ClarkMeyer} that the threshold strain for the onset of buckling is
\begin{eqnarray}
\alpha_T = \frac{2\pi}{L_z}\left( \frac{K_B}{B} \right)^{1/2}\;,
\label{Buckling Threshold Strain}
\end{eqnarray}
where $L_z$ is the height of the cell, measured along the layer normal, $B$ is the smectic bulk modulus and $K_B$ is the smectic bend modulus. Using typical values ($L_z=10^{-4}$ m, $B=10^7$ N/m$^2$, $K_B=10^{-11}$ N) one finds $\alpha_T \sim 10^{-4}$.

In the above cases non-zero strain is introduced by forcing the layer spacing to deviate from its preferred value $d_p$ by moving the top and bottom plates closer together or further apart, thus changing $d$. Another way to introduce strain is to fix the top and bottom plates (thus fixing the imposed layer spacing $d$) but to change the preferred spacing $d_p$ by having the molecules tilt, e.g., through the bulk electroclinic effect \cite{SelingerBECE}. It can be seen from Fig.~\ref{smectic cartoon} that the preferred layer spacing $d_p$ is given by:
\begin{eqnarray}
d_p=d_A \cos(R \theta_p)\;,
\label{LayerContractionRelation}
\end{eqnarray}
where $\theta_p$ is the preferred tilt and $d_A$ is the layer spacing of the Sm-$A$ phase, i.e., in the absence of tilt. The reduction factor $R$ is a measure of the de Vries-like nature of the smectic \cite{LemieuxR} . For an ideal de Vries smectic, which exhibits no change in layer spacing at the transition, $R=0$, while for a traditional transition, shown in Fig.~\ref{smectic cartoon}, $R\approx 1$. Typical de Vries smectics have $0.2 < R < 0.5$.

\begin{figure}
\begin{center}
\includegraphics[scale=0.7]{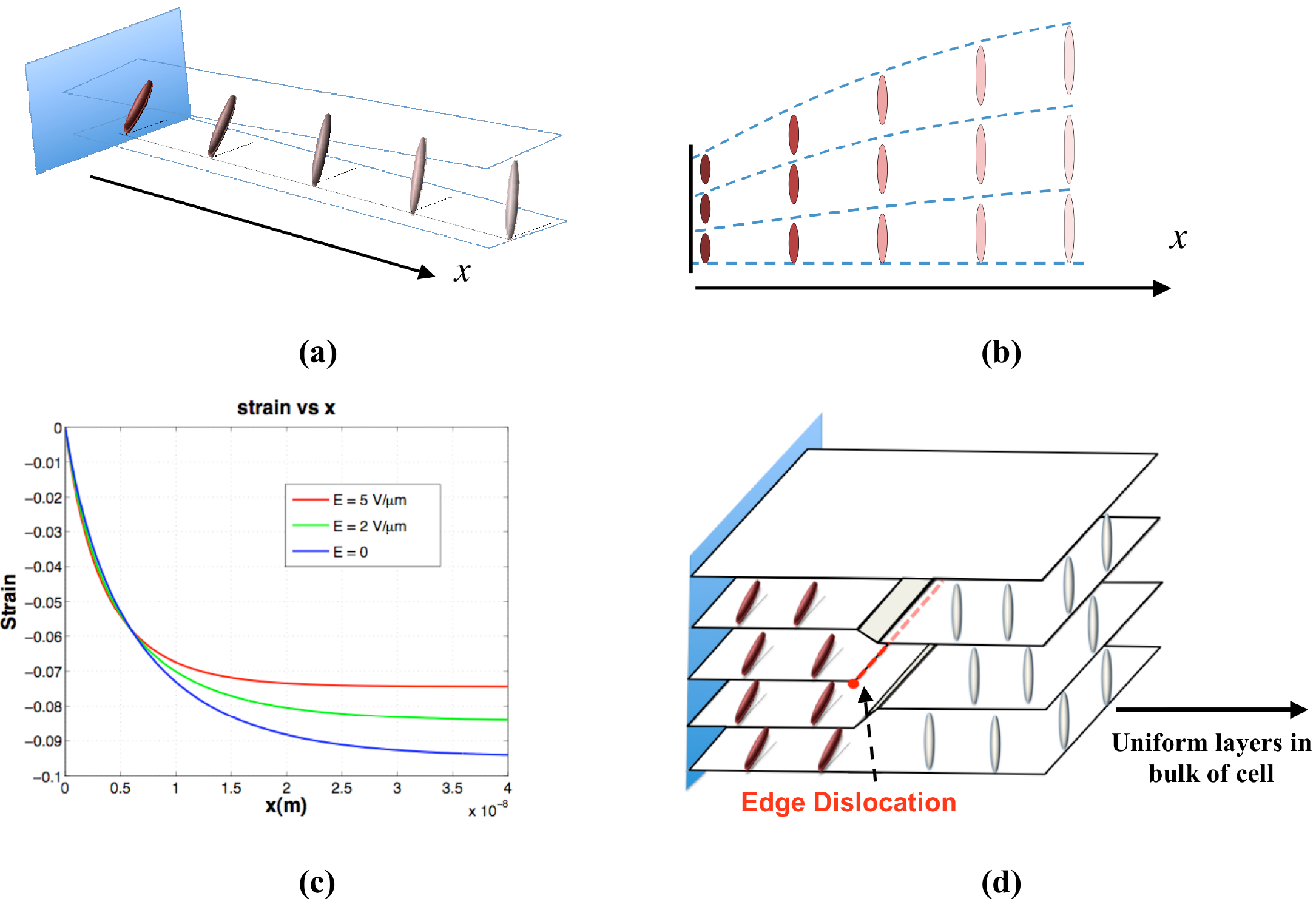}
\caption{(a) A single layer schematic view of tilt and layer spacing variation for $|\theta_S| > |\theta_B|$. (b) A side view showing multiple layers. As one moves further into the bulk the magnitude of compressive strain grows. (c) A plot of strain vs position $x$ for a variety of fields $E>0$. (d) Compressive strain is relieved through the removal of layers, whereby the bulk of the cell has a uniform layer spacing.}
\label{CompressiveStrainSchematic}
\end{center}
\end{figure}

We now consider the effects of strain for a chiral system that has both surface and bulk tilt. For simplicity we first consider a system with a continuous Sm-$A^*$--Sm-$C^*$ transition and with de Vries reduction factor $R=1$.  From Fig.~\ref{thetaSvsEcontAC} it can be seen that at any field the surface tilt $\theta_S$ and the bulk tilt $\theta_B$ will be different. This means that preferred tilt $\theta_p$ will vary continuously as one moves between the surface and bulk of the cell. Correspondingly, the preferred layer spacing $d_p(x)=d_A\cos(\theta(x))$ is also position dependent. Upon cooling into the Sm-$A^*$ phase, the layers first nucleate at the surface and then grow into the bulk \cite{MaclennanFieldControl} so the imposed layer spacing is $d=d_S=d_A\cos(\theta_S)$. Therefore the corresponding position dependent strain $\alpha(x)$ is:
\begin{eqnarray}
\alpha(x)= 1 - \frac{d_p(x)}{d_S} = 1 -  \frac{\cos \left(\theta(x)\right)}{\cos(\theta_S)}\;, 
\label{Position Dependent Strain}
\end{eqnarray}
where $\theta(x)$, can be obtained by integrating Eq.~(\ref{First integral}). For zero applied field (and thus zero bulk tilt) $\theta(x)$ decreases and $d_p(x)$ increases as one move from surface to bulk, as shown in Figs.~\ref{CompressiveStrainSchematic}(a) and (b). The corresponding compressive strain vs position $x$ is shown in Fig.~\ref{CompressiveStrainSchematic}(c) and one can see that the magnitude of the strain is largest in the bulk. If this maximum strain exceeds the threshold for dislocation formation then layers will be removed from the bulk, and the bulk of the cell will have a uniform layer spacing $d_B=d_A \cos(\theta_B)$. For $E>0$ the magnitude of bulk tilt $|\theta_B(E)|\neq0$ but is still smaller than $|\theta_S(E)|$ so the strain is still compressive. Sample plots of strain vs $x$ for various values of $E>0$ are shown in Fig.~\ref{CompressiveStrainSchematic}(c). The maximum strain magnitude for non-zero $E$ is smaller than that of zero $E$ because as $E$ becomes increasingly negative, the difference between bulk and surface tilts decreases. 

For $E<-|E_0|$, where $|E_0|$ is the magnitude of $E$ that eliminates surface tilt, the situation is reversed. As shown in Figs.~{\ref{DilativeStrainSchematic}(a) and (b), the tilt is now largest in the bulk and smallest at the surface. Thus the strain is dilative throughout the cell. Figure~\ref{DilativeStrainSchematic}(c) shows sample plots of strain vs $x$ for a variety of fields $E<-|E_0|$. Once again the strain increases as one moves into the cell. Once $\alpha(x)$ exceeds $\alpha_T \sim 10^{-4} $ the layers will buckle in order to increase the average layer spacing, as shown in Fig.~\ref{DilativeStrainSchematic}(d). From Fig.~\ref{DilativeStrainSchematic}(c) it can be seen that the typical maximum strain is much larger than $\alpha_T$ and is reached for  $x \gtrsim \xi$, where $\xi$ is the correlation length, which is much smaller than the width $L_x$ of the cell. Thus buckling will occur in the majority of the cell which may explain the experimental observation \cite{MaclennanFieldControl} that for $E<-|E_0|$ the layer orientation becomes increasingly inhomogeneous, which diminishes the optical properties of the cell.

\begin{figure}
\begin{center}
\includegraphics[scale=0.7]{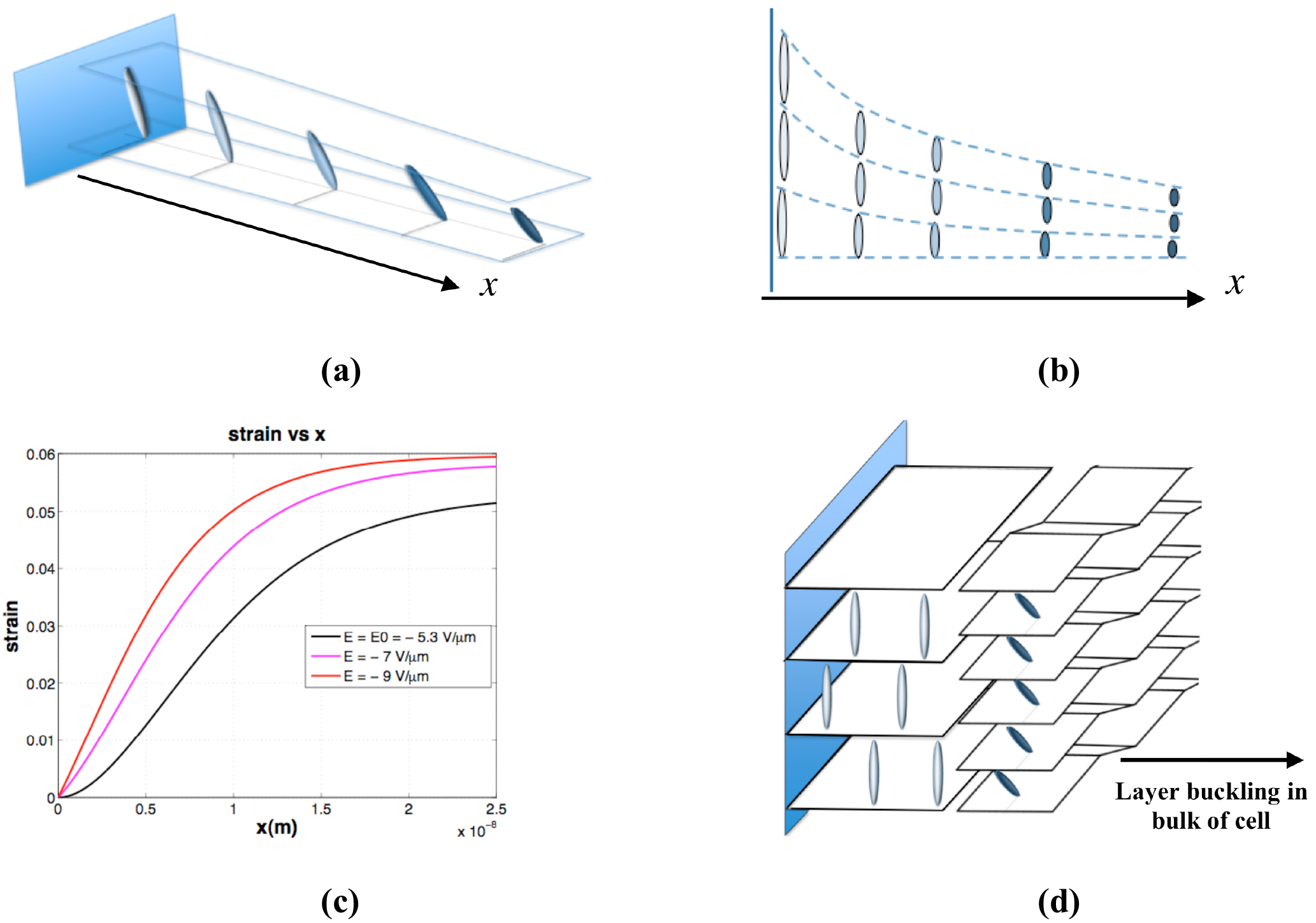}
\caption{(a) A single layer schematic view of tilt and layer spacing variation for $|\theta_S| < |\theta_B|$. (b) A side view showing multiple layers. As one moves further into the bulk the magnitude of dilative strain grows. (c) A plot of strain vs position $x$ for a variety of fields $E\leq -|E_0|$. (d) Dilative strain is most easily relieved through the buckling of the layers throughout the bulk of the cell. }
\label{DilativeStrainSchematic}
\end{center}
\end{figure}

For $-|E_0|<E<0$ the signs of $\theta_S$ and $\theta_B$ differ so $\theta(x)$ must change sign as one moves into the bulk. This situation, shown schematically in Fig.~\ref{CompressiveAndDilativeStrainSchematic}(a) and (b), is significantly more complicated because there may be a combination of compressive {\it and} dilative strain.  If there is compressive strain at small $x$ followed by dilative strain at larger $x$ then one may have dislocations only, as shown in Fig.~\ref{CompressiveAndDilativeStrainSchematic}(c), or one may have a combination of dislocations {\it and} buckling, as shown in Fig.~\ref{CompressiveAndDilativeStrainSchematic}(d). Determining which is the energetically preferred scenario is beyond the scope of this article but based on the experimental work \cite{MaclennanFieldControl} we suspect it is the former, i.e., dislocations only. As discussed above, a significant drop of in cell quality is observed for $E<-|E_0|$, which we believe to be a result of buckling due to dilative strain. If, in the regime $-|E_0|<E<0$, strain was relieved by dislocations {\it and} buckling, then we would expect to see a drop of in cell quality at a field smaller than $-|E_0|$. 

\begin{figure}
\begin{center}
\includegraphics[scale=0.7]{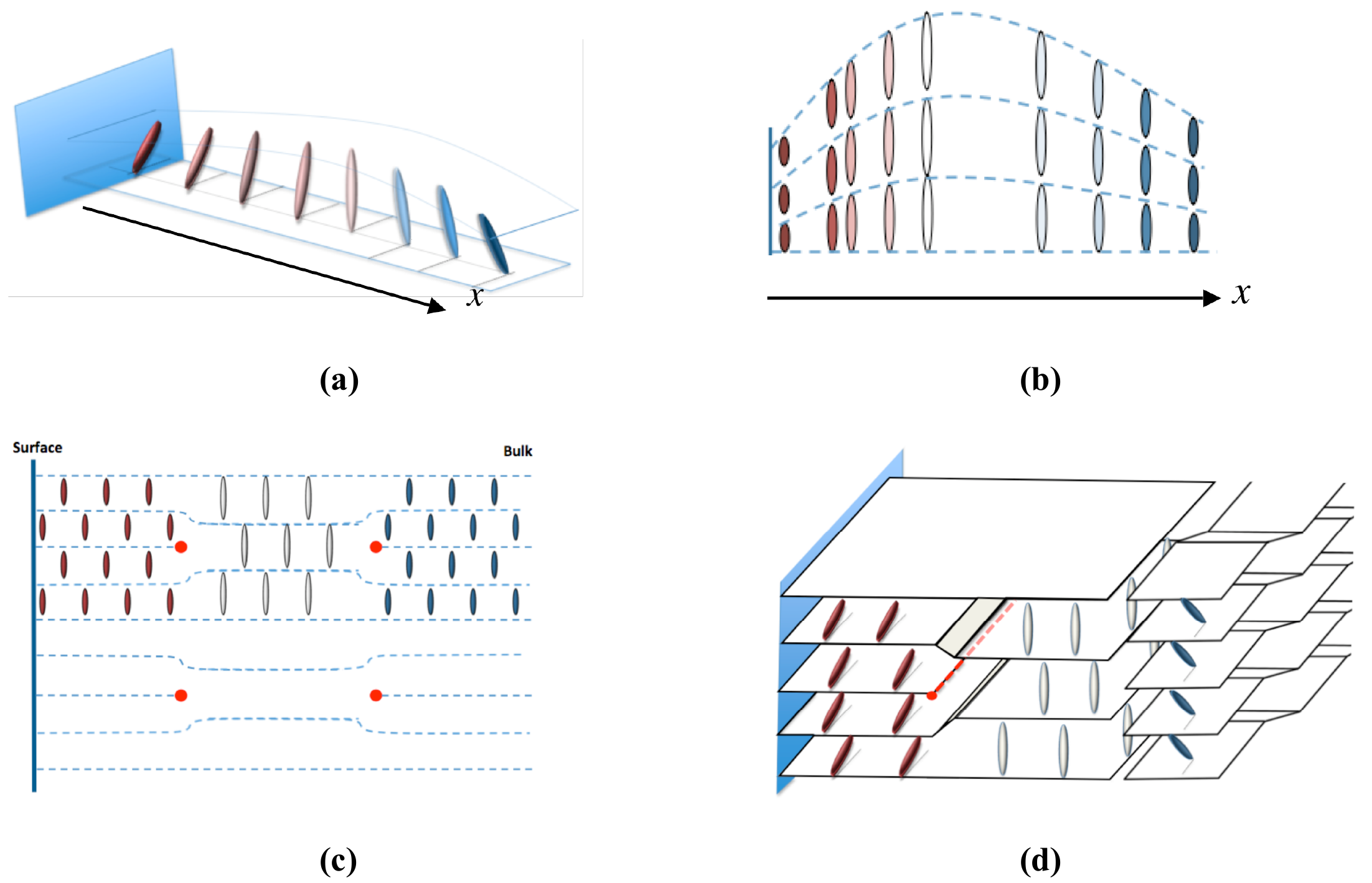}
\caption{(a) A single layer schematic view of tilt and layer spacing variation when $\theta_S$ and $\theta_B$ have different signs. (b) A side view showing multiple layers. If there is compressive strain at small $x$ followed by dilative strain at larger $x$, one may have dislocations only, as shown in (c), or one may have a combination of dislocations {\it and} buckling, as shown in (d).}
\label{CompressiveAndDilativeStrainSchematic}
\end{center}
\end{figure}

The threshold field for the onset of buckling should be larger for de Vries smectics which have small tilt induced layer contraction. One should however, keep in mind that the de Vries materials typically have transitions at or near tricriticality. As shown in Figs.~\ref{BECE_Response_curves} and \ref{SECE_Response_curves}, the closer to tricriticality, the larger the surface and bulk electroclinic effects. This means that the difference between surface and bulk tilt (and hence layer spacings) will be larger, the effect of which is to increase the strain. Thus in de Vries materials the effect of a smaller reduction factor $R$ may be offset by the larger induced tilts due to proximity to tricriticality. 

For materials with a first order Sm-$A^*$--Sm-$C^*$ transition we expect the strain effects to be qualitatively similar to those for a material with a continuous Sm-$A^*$--Sm-$C^*$ transition. For ranges of $E$ in which the surface tilt is significantly different from bulk tilt, as shown in Fig.~\ref{E0_vs_tau_1stAC}(c), the maximum strain will be higher. However, despite the discontinuity  between surface and bulk tilt, the tilt profile $\theta(x)$ can be shown to be continuous. Thus, as in the zero field case \cite{Rudquist&Saunders}, $\theta(x)$ and $\alpha(x)$ may each display a kink, but not a  discontinuity. We still expect the onset of buckling to occur for fields beyond which the surface tilt is reversed, i.e., for fields smaller than $-|E_0|$

It should be pointed out that the above analysis only considers the effect of the tilt on layer spacing (and thus strain) but not vice versa. In effect, we have assumed that the system energetics are dominated by the tilt, with layer spacing effects being secondary. A full analysis should treat the layer spacing and the tilt on an equal energetic footing. One would, for example, expect that the strain could also be relieved by reducing the difference between bulk and surface tilt. This would presumably increase the threshold field above which layer buckling would occur. Such a full analysis, treating {\it both} tilt and layer spacing on an equal footing, is beyond the scope of this article. In fact, such an analysis has not even been carried out for the case of the purely bulk electroclinic effect \cite{BECE_only_buckling}. However, despite its shortcomings, we believe that the above analysis provides a useful qualitative framework to understand the role played by layering in the field control of the surface electroclinic effect. In particular, our conclusion that the drop-off in cell quality for $E<-|E_0|$ is due to the onset bucking should remain valid. Of course, we would welcome further experimental investigation to test this hypothesis.

\section{Summary}
\label{Summary}

In summary, we have presented and analyzed a model for the combined bulk and surface electroclinic effects in the Sm-$A^*$ phase near a Sm-$A^*$--Sm-$C^*$ transition that is either continuous or first order. In particular, we have studied how the surface induced tilt can be controlled by an external field, and have shown that it can be reversed or even eliminated. Our predictions for the field control of the surface tilt near a continuous Sm-$A^*$--Sm-$C^*$ transition are in good agreement with experiments on W415 \cite{MaclennanFieldControl}. Our model also allows us to calculate the variation of $|E_0|$ (the field magnitude required to eliminate surface tilt) with $\tau$ (the width of the Sm-$A^*$--Sm-$C^*$ phase temperature window). We show that $|E_0|(\tau)$ is monotonically increasing which, perhaps surprisingly, means that a smaller $|E_0|$ is required to eliminate  larger zero-field surface tilt.

We also present, for the first time, an analysis of surface tilt field control near a first order Sm-$A^*$--Sm-$C^*$ transition. Depending on the proximity of the transition, the surface tilt can be continuously varying with $E$ or (if sufficiently close to the transition) it can exhibit discontinuities and hysteresis as $E$ is varied. As with field control of the surface tilt near a continuous Sm-$A^*$--Sm-$C^*$ transition we find that the surface tilt can be reversed or eliminated at field magnitude $|E_0|$. The dependence of $|E_0|$ on proximity to the first order Sm-$A^*$--Sm-$C^*$ transition is qualitatively similar to that for the continuous transition. 

As discussed throughout this article, because the layers are fixed upon formation at the transition to the Sm-$A^*$ phase, the proximity to the Sm-$A^*$--Sm-$C^*$ transition is also effectively fixed. One could perhaps quench the system into the Sm-$A^*$ phase at lower temperatures, or use different chemical homologs to vary proximity to the Sm-$A^*$--Sm-$C^*$ transition, but the easiest way may be to vary the enantiomeric excess of the system. We discuss this in some detail, showing that increasing the enantiomeric excess will effectively narrow the temperature range of the Sm-$A^*$ phase, thus bringing the system closer to the Sm-$A^*$--Sm-$C^*$ transition. However, one must be careful to also account for the fact that increasing the enantiomeric excess will also increase the effective coupling between the tilt and both the applied field and the effective surface field. As part of this analysis we show how the field dependence of surface tilt varies with enantiomeric excess, near both continuous and first order Sm-$A^*$--Sm-$C^*$ transitions. For a first order transition, the effect of increasing enantiomeric excess can be dramatic, going from continuous to discontinuous variation of surface tilt with applied field . 

Lastly, we analyze the effect that the surface and bulk tilt has on the layers. The difference in surface and bulk tilts means that there will be a difference between surface and bulk layer spacings. This in turn means that there will be a non-zero, position dependent strain throughout the liquid crystal cell. For fields such that the magnitude of surface tilt is larger than the magnitude of bulk tilt, the strain will be compressive and will eventually be relieved by dislocations, i.e., the removal of layers in the bulk. This results in a cell whose bulk layering is uniform. For fields such that the surface tilt is reversed (from its original zero field direction), we show that the strain is dilative. In this case the strain is most easily relieved through buckling throughout the bulk of the cell. This may explain the experimental observation \cite{MaclennanFieldControl} that for for such fields the layer orientation becomes increasingly inhomogeneous and diminishes the optical properties of the cell. We also discuss the situation for de Vries smectics, whose tilt induced change in layer spacing is small. Thus one would expect the corresponding strain to also be small. However, one must be careful because de Vries materials typically have transitions at or near tricriticality which makes surface and bulk electroclinic effects larger. Thus in de Vries materials the effect of a smaller tilt induced change in layer spacing may be offset by the larger induced tilts due to proximity to tricritcality. 

We hope that  the many new predictions we have made in this article, in particular for materials with a first order Sm-$A^*$--Sm-$C^*$ transition, motivate further experimental investigation of the field dependence of surface tilt in the near future.

K. Z. , D. N. H. and K. S. acknowledge support from the National Science Foundation under Grant No. DMR-1005834.


\begin{thebibliography}{99}

%1
\bibitem{LemieuxR} J. C. Roberts, N. Kapernaum, Q. Song, D. Nonnemacher, K. Ayub, F. Giesselmann and R. P. Lemieux, J. Am. Chem. Soc. {\bf 132}, 364 (2010).

%2
\bibitem{Meyer} R. B. Meyer, Mol. Liq. Crys. {\bf 40}, 33 (1977).

%3
\bibitem{Garoff and Meyer} S. Garoff and R. B. Meyer, Phys. Rev. Lett. {\bf 38}, 848 (1977).

%4
\bibitem{Huang&Viner} C. C. Huang and J. M. Viner, Phys. Rev. A {\bf 25}, 3385 (1982).

%5
\bibitem{HuangTP} H.Y. Liu, C. C. Huang, Ch. Bahr and G. Heppke, Phys. Rev. Lett. {\bf 61}, 345 (1988).

%6
\bibitem{ShashidharTP} R. Shashidhar, B. R. Ratna, Geetha G. Nair, S. Krishna Prasad, Ch. Bahr and G. Heppke, Phys. Rev. Lett. {\bf 61}, 547 (1988).

%7
\bibitem{Bahr&Heppke1}  Ch. Bahr and G. Heppke, Phys. Rev. A {\bf 39}, 5459 (1989).

%8
\bibitem{Bahr&Heppke2}  Ch. Bahr and G. Heppke, Phys. Rev. A {\bf 41}, 4335 (1990).

%9
\bibitem{Xue&Clark}  J. H. Xue and N. A. Clark, Phys. Rev. Lett. {\bf 64}, 307 (1990).

%10
\bibitem{Chen&Ouchi}  W. Chen, Y. Ouchi, T. Moses, and Y. R. Shen and K. H. Yang, Phys. Rev. Lett. {\bf 68}, 1547 (1992).

%11
\bibitem{Rovsek&Zeks}  B. Rovsek and B. Zeks, Mol. Cryst. Liq. Cryst.  {\bf 263}, 49 (1995).

%12
\bibitem{Shao&Boulder} R.-F. Shao, J. E. Maclennan, N. A. Clark, D. J. Dyer, D. M. Walba, Liq. Cryst. {\bf 28}, 117, (2001). 

%13
\bibitem{SECE with Varying T} Another approach would be to establish an alignment direction for ${\bf \hat N}$ without doing so for ${\bf \hat n}$, i.e., align the smectic layers in a homogeneous bookshelf structure without having a rubbing direction at the surfaces. This would then allow ${\bf \hat n}$ to rotate, and thus $\theta_S$ to change, at the surface as temperature is varied. One way to do this is to shear-align the sample, i.e. to slide the top substrate with respect to the lower substrate under an applied a.c. field, as was done with the first experiments on surface-stabilized ferroelectric liquid crystal cells. See: N. A. Clark, S. T. Lagerwall, Appl. Phys. Lett. 36, 899 (1980); N. A. Clark, M. A. Handschy, S. T. Lagerwall, Mol. Cryst. Liq. Cryst. {\bf 94}, 213 (1983).

%14
\bibitem{Rudquist&Saunders} K. Saunders and P. Rudquist, Phys. Rev. B {\bf 83}, 051711 (2011).

%15
\bibitem{415Footnote} It is possible that W415 has a Sm-$A^*$ -- Sm-$C^*$ that is tricritical or very weakly first order. Unfortunately, the nature of the bulk transition is not discussed in the article \cite{Shao&Boulder}.  However, that the transition is not first order can be inferred from the lack of an S-shaped character in the tilt versus field curves for the bulk electroclinic effect (Fig. 4 in \cite{Shao&Boulder}).

%16
\bibitem{MaclennanFieldControl} J. E. Maclennan, D. Muller, R.-F. Shao, D. Coleman, D. J. Dyer, D. M. Walba and N. A. Clark , Phys. Rev. E. {\bf 69}, 061716 (2004)

%17
\bibitem{racemic first order transition} We consider a racemic bulk Sm-$A$ - Sm-$C$ transition that is first order, i.e., $u<0$. It should be pointed out that the transition can also be driven first order by increasing the enantiomeric excess \cite{Liu&Huang&Min}. For the sake of simplicity, we do not consider that possibility here.

%18
\bibitem{Liu&Huang&Min}  H. Y. Liu, C. C. Huang, T. Min, M. D. Wand, D. M. Walba, N. A. Clark, Ch. Bahr and G. Heppke, Phys. Rev. A {\bf 40}, 6759 (1989).

%19
\bibitem{e and EE} How $e$ varies with enantiomeric excess is not obvious. For low enantiomeric excess (assumed here) the variation will be approximately linear. However, given that $e$ must plateau as enantiomeric excess reaches saturation, $e$ must vary nonlinearly with large enantiomeric excess.

%20
\bibitem{ClarkMeyer} N. A. Clark and R. B. Meyer, Appl. Phys. Lett. {\bf 22}, 493 (1973).

%21
\bibitem{SelingerBECE} R. E. Geer, S. J. Singer, J. V. Selinger, B. R. Ratna, and R. Shashidhar, Phys. Rev. E, {\bf 57}, 3059 (1998).

%22
\bibitem{BECE_only_buckling} The onset of buckling in a fixed length cell via bulk electroclinic reduction of layer spacing has been modeled \cite{SelingerBECE}. This model did not consider surface effects. As with our analysis, this model only considered the effect of the tilt on layer spacing (and thus strain) and not vice versa. We have carried out a preliminary analysis \cite{SaundersUnpublished} of the BECE-only system, in which we treat layer spacing and tilt on an equal energetic footing. Our analysis shows that the threshold field for the onset of buckling is indeed increased, when one allows layering strain to affect tilt. 

%23
\bibitem{SaundersUnpublished} K. Saunders, to be published.

\end{thebibliography}
\end{document}